\documentclass[a4paper]{article}
\usepackage{pdfsync}
\usepackage[utf8]{inputenc}
\usepackage{graphicx}
\usepackage{amssymb,amsfonts,amsmath,amsthm}
\usepackage{hyperref}
\usepackage{graphicx}
\usepackage{multirow}
\usepackage{verbatim}
\usepackage{color}
\usepackage{bm}
\usepackage[sectionbib,round]{natbib}
\usepackage{orcidlink}

\theoremstyle{plain}

\newtheorem{proposition}{Proposition}[section]
\theoremstyle{definition}

\newtheorem{remark}{Remark}[section]

\begin{document}

\title{\textbf{Subarchetypal Analysis: A Semi-Supervised Framework for Archetypal Analysis}}

\author{Vicente J. Bol\'os$^{1}$\,\orcidlink{0000-0002-9099-8200}, Mireia Mollar-Gumbau$^2$\,\orcidlink{0009-0005-6250-4970}, Aleix Alcacer$^2$\,\orcidlink{0000-0003-3566-0204},\\ Irene Epifanio$^{2,3,*}$\,\orcidlink{0000-0002-6973-311X} \\
{\scriptsize $^1$ Dept. Business Mathematics, Universitat de Val\`encia,
Av. Naranjos s/n, Val\`encia, 46022, Spain.} \\
{\scriptsize $^2$ Dept. Mathematics, Universitat Jaume I, Campus Riu Sec, 12071, Castell\'o de la Plana, Spain.} \\
{\scriptsize $^3$ ValgrAI - Valencian Graduate School and Research Network of Artificial Intelligence.} \\
{\scriptsize e-mail\textup{: \texttt{vicente.bolos@uv.es} (V. J. Bol\'os), \texttt{mollarm@uji.es} (M. Mollar-Gumbau),}} \\
{\scriptsize \textup{ \texttt{aalcacer@uji.es} (A. Alcacer)}, \texttt{epifanio@uji.es} (I. Epifanio)} 
 \\}

\date{}

\maketitle

\begin{abstract}
In this work, we introduce \textit{subarchetypal analysis}, a new semi-supervised framework for Archetype Analysis (AA) and Archetypoid Analysis (ADA) based on subsets of observations from which archetypal profiles are generated. To the best of our knowledge, this is the first subset-constrained framework for archetypal analysis. Classical AA and ADA are recovered as special cases, and the location result for archetypes is extended to the proposed framework. The generating subsets may be defined \textit{a priori} by the user to incorporate external knowledge, or determined from intrinsic properties of the data or the underlying problem structure. As an illustration of the latter, we consider \textit{efficient subarchetypes} within Data Envelopment Analysis, where the subsets are determined by the geometry of the efficient frontier. We develop and implement algorithms for subarchetypal analysis in both multivariate and functional settings, available in an \textsf{R} package. The proposed methods are compared with constrained clustering, group-wise archetypal analysis, and classical unconstrained AA and ADA, and their practical usefulness is demonstrated through applications to business, sports analytics, and human development data.
\end{abstract}


\section{Introduction}

Archetype Analysis (AA) \citep{cutler1994archetypal} is an unsupervised learning technique that seeks to identify extremal patterns in datasets. In AA, each observation is approximated as a convex combination of a set of archetypes, which are themselves convex combinations of the observed data points. As a result, archetypes constitute representative profiles that allow us to interpret the variability of the dataset. This formulation provides a powerful framework for dimensionality reduction and feature extraction, as comprehensively reviewed by \citet{alcacer2025survey}. Archetypoid Analysis (ADA) \citep{vinue2015archetypoids} extends AA by restricting the representatives to observed data points rather than convex combinations of observations, while preserving the convex representation of the observations. This makes the resulting representatives easier to interpret in applications where actual data instances are preferred. Although AA and ADA were originally developed for multivariate numerical data, they have been successfully extended to functional data \citep{Epifanio2016}.

Throughout the paper, the term ``archetypal analysis'' is used to refer to the family of archetype-based methods, including AA, ADA, and their extensions, such as the corresponding functional methodologies. 
Archetypal analysis lies conceptually between principal component analysis and clustering techniques, providing a low-dimensional representation of the data while retaining an interpretation in terms of representative archetypal profiles \citep{morup2012archetypal}.

To date, archetypal analysis has been developed exclusively in an unsupervised setting, with no established semi-supervised framework for exploiting expert knowledge through constraints on the archetypal construction process. As in semi-supervised clustering \citep{GonzalezAlmagro2025}, archetypal analysis can benefit from the incorporation of expert-defined constraints when prior knowledge is available, thereby guiding the identification of archetypal profiles toward more meaningful and interpretable representations. To address this gap, we introduce, for the first time, a semi-supervised archetypal analysis framework based on subset constraints on the construction of archetypal profiles, which we refer to as \textit{subarchetypal analysis}.

Semi-supervised learning is the branch of machine learning that focuses on using both labelled and unlabelled data to perform specific learning tasks \citep{van2020survey}. 
While the vast majority of studies on semi-supervised learning concentrate on semi-supervised classification, the present work diverges from this standard task. Our approach shares a conceptual affinity with semi-supervised clustering \citep{jones2022discriminative}, as prior labelled information is embedded directly into the data organizing process rather than used to learn a mapping function. However, unlike traditional semi-supervised clustering (which strictly focuses on utilizing group labels to partition the entire dataset), our framework applies these labels specifically to find the extreme representatives of the data. Specifically, this framework operates by injecting known labels as structural constraints to govern subarchetypal analysis, leveraging the prior group information to explicitly restrict and dictate the admissible subsets within which the archetypes themselves can be constructed.

Possible applications of this semi-supervised archetypal analysis arise in settings where partial expert knowledge is available and can be encoded as constraints on the archetypal structure. For instance, in biomedical data analysis, clinicians may impose constraints reflecting known phenotypes or disease subgroups to guide the discovery of meaningful archetypal patient profiles. In market segmentation, partial labelling of customer types can be used to steer archetypal profiles toward interpretable consumer archetypes consistent with business knowledge. Similarly, in remote sensing and environmental science, expert knowledge about land-cover classes or ecological regions can be incorporated to ensure that the extracted archetypes respect known physical or geographical structures. In sports analytics, prior information about player roles or tactical positions can be used to constrain the learning process so that the resulting archetypes correspond to interpretable playing styles or strategic profiles. In human development studies, socio-economic or demographic constraints can be incorporated to ensure that the resulting archetypal representatives reflect meaningful population profiles consistent with known development patterns and inequalities.

In some applications, such constraints may arise from the intrinsic structural properties of the problem rather than from expert knowledge, yielding a data-driven setting. A notable example is provided by Data Envelopment Analysis (DEA) \citep{Charnes1978}, where the geometry of the \textit{efficient frontier} naturally induces restrictions on the admissible archetypes by requiring them to be efficient.
The motivation for the present work is rooted precisely in the computation of such \textit{efficient subarchetypes}.

The main contributions of this work are as follows: (i) we introduce, to the best of our knowledge, the first subset-constrained framework for AA and ADA, in which archetypal profiles are generated from user-defined or data-driven subsets. This framework includes classical AA and ADA as special cases and extends the location result for archetypes; (ii) we develop algorithms for both the multivariate and functional settings and implement them in the \texttt{subarchetypes} package for \textsf{R} \citep{R26}; (iii) we show that the proposed framework naturally accommodates the computation of efficient subarchetypes in DEA, where the generating subsets are determined by the facets of the efficient frontier; and (iv) we demonstrate its practical usefulness through comparisons with alternative approaches and applications to multivariate and functional human development data.

The remainder of the paper is organized as follows. Section~\ref{sec:MB} provides the necessary mathematical background. In particular, Section~\ref{sec:AA} reviews the fundamentals of archetypal analysis, whereas Section~\ref{sec:DEA} introduces the basic concepts of DEA. Section~\ref{sec:method} introduces the proposed semi-supervised archetypal analysis framework and extends the location result for archetypes. Section~\ref{sec:algo} develops the corresponding algorithms, including a simulation study of their computational cost and heuristic strategies for screening candidate configurations. Section~\ref{sec:results} presents motivating applications of the proposed methodology, including efficient subarchetypes in DEA and multivariate and functional applications to human development data.
Finally, Section~\ref{sec:conc} concludes the paper and outlines directions for future research. The proofs are provided in Appendix~\ref{sec:proofs}. Additional illustrative comparisons with constrained clustering (a form of semi-supervised clustering), group-wise archetypal analysis, and classical unconstrained archetypal analysis are reported in Appendix~\ref{sec:ilcomp}.

\section{Mathematical background}
\label{sec:MB}

\subsection{Archetypal analysis}
\label{sec:AA}

This section provides the basic background on archetypal analysis required for the remainder of the paper. We first review the concepts of archetypes and archetypoids for multivariate observations and then present their extension to functional observations through functional archetypal analysis.

\subsubsection{Archetypes and archetypoids}
\label{sec:AAADA}

Let $\left\{ \boldsymbol{x}_i\right\} _{i=1}^n$ denote a sample of $m$-dimensional observations.
In Archetype Analysis (AA), the $k$ archetypes $\left\{ \boldsymbol{z}_j\right\} _{j=1}^k$ are defined as convex combinations of the observations:
\begin{equation}
\label{eq:arch}
\boldsymbol{z}_j = \sum_{l=1}^n \beta_{jl} \boldsymbol{x}_l,\qquad j=1,\ldots,k,
\end{equation}
with $\beta_{jl} \geq 0$ and $\sum_{l=1}^n \beta_{jl} = 1$. In matrix form, \eqref{eq:arch} is expressed as $\boldsymbol{Z}_{k\times m} = \boldsymbol{\beta}_{k\times n}\boldsymbol{X}_{n\times m}$.
Each observation is then approximated as a convex combination of the archetypes:
\begin{equation}
\label{eq:alpha}
\boldsymbol{x}_i \approx \hat{\boldsymbol{x}}_i = \sum_{j=1}^k \alpha_{ij} \boldsymbol{z}_j,\qquad i=1,\ldots,n,
\end{equation}
with $\alpha_{ij} \geq 0$ and $\sum_{j=1}^k \alpha_{ij} = 1$, which in matrix form is given by $\hat{\boldsymbol{X}}_{n\times m}=\boldsymbol{\alpha}_{n\times k}\boldsymbol{Z}_{k\times m}$.
Under these constraints, we minimize the residual sum of squares
\begin{equation}
\label{eq:rss}
\begin{split}
\operatorname{RSS}
&= \sum_{i=1}^n \left\lVert \boldsymbol{x}_i - \hat{\boldsymbol{x}}_i\right\rVert ^2 = \sum_{i=1}^n \left\lVert \boldsymbol{x}_i - \sum_{j=1}^k \alpha_{ij} \sum_{l=1}^n \beta_{jl} \boldsymbol{x}_l
\right\rVert ^2 \\
&= \sum_{i=1}^n \sum_{g=1}^m
\left(
x_{ig}
- \sum_{j=1}^k
\alpha_{ij}
\left(
\sum_{l=1}^n
\beta_{jl}\, x_{lg}
\right) 
\right) ^2 ,
\end{split}
\end{equation}
where $\| \cdot \|$ represents the Euclidean norm. In certain cases, when the variables corresponding to each observation differ in nature and are expressed in different units of measurement, the data should be brought to a common scale prior to analysis, for example by standardization, normalization, or rank transformation.

In summary, parameters $\beta _{jl}$ define archetypes from observations, 
while $\alpha_{ij}$ describe how each observation is approximated as a mixture of these archetypes. 

In Archetypoid Analysis (ADA) \citep{vinue2015archetypoids}, the $k$ archetypoids $\left\{ \boldsymbol{z}_j\right\} _{j=1}^k$ are defined as single observations. So, they are given by \eqref{eq:arch} with $\beta_{jl} \in \left\{ 0,1\right\} $ and $\sum_{l=1}^n \beta_{jl} = 1$. Each observation is also approximated as a convex combination of the archetypoids in the same way as \eqref{eq:alpha}, and the residual sum of squares given by \eqref{eq:rss} is minimized.

The number of archetypes or archetypoids $k$ may be chosen by a domain expert based on prior knowledge or determined using an elbow-type criterion,  the most widely used approach in practice, selecting the value of $k$ at which the $\operatorname{RSS}$ curve flattens \citep{cutler1994archetypal,eugster2009spider,vinue2015archetypoids}. Alternatively, other criteria may be adopted, such as the model consistency criterion proposed by \citet{hinrich2016archetypal}.

\subsubsection{Functional archetypal analysis} 
\label{sec:faa}

Let $\{x_i(t)\}_{i=1}^n$ be a set of univariate observed functions defined on the interval $[a,b]$. Throughout this work, we assume that the functions satisfy suitable smoothness conditions and belong to the Hilbert space $L^2[a,b]$ of square-integrable functions. The objectives of Functional Archetype Analysis (FAA) and Functional Archetypoid Analysis (FADA) are the same as in the multivariate case, replacing vectors by functions and the Euclidean norm by the $L^2$-norm.

Using a basis representation, each function is expressed as $x_i(t)=\boldsymbol{b}_i^\prime\boldsymbol{B}$, where $\boldsymbol{b}_i$ is the vector of basis coefficients of the $i$-th observed function, $\boldsymbol{B}$ is the vector of basis functions, and $^\prime$ denotes transpose. The $\operatorname{RSS}$ can then be written as \citep{Epifanio2016}
\begin{equation}
\label{RSSfar}
\operatorname{RSS}=\sum_{i=1}^n \boldsymbol{a}_i^\prime
\boldsymbol{W}\boldsymbol{a}_i,
\end{equation}
where $\boldsymbol{a}_i^\prime=\boldsymbol{b}_i^\prime-\sum_{j=1}^k\alpha_{ij}\sum_{l=1}^n\beta_{jl}\boldsymbol{b}_l^\prime$, and $\boldsymbol{W}$ is the matrix of inner products between basis functions. For an orthonormal basis, $\boldsymbol{W}$ is the identity matrix, reducing FAA (FADA) to AA (ADA) applied to the basis coefficients.

\subsection{Data envelopment analysis}
\label{sec:DEA}

In this section, we review the theoretical framework and fundamental concepts of Data Envelopment Analysis (DEA) that are required to develop the notion of efficient subarchetypes in Section~\ref{sec:efsuba}.
To provide the necessary context, DEA is a non-parametric methodology for assessing the relative efficiency of a homogeneous set of Decision-Making Units (DMUs) that transform multiple inputs into multiple outputs \citep{Cooper2007}. Through mathematical programming models, DEA identifies the best-performing DMUs and constructs the corresponding efficient frontier. Units located on this frontier are classified as efficient, whereas those outside it are deemed inefficient. %The growing body of recent work based on DEA underscores its wide applicability and continued relevance in solving practical optimization and efficiency assessment problems \citep{chen2024controlled,amirteimoori2024improving}.

\subsubsection{Efficient frontier}
\label{sec:efr}

Let $\mathcal{D}=\left\{ \textrm{DMU}_1, \ldots ,\textrm{DMU}_n \right\} $ be a set of $n$ DMUs with $m_1$ inputs and $m_2$ outputs. Let us denote by $x_{ig_1}$ and $y_{ig_2}$ the $g_1$-th input and $g_2$-th output, respectively, of the $i$-th DMU. We also assume that $x_{ig_1}$ and $y_{ig_2}$ are all positive, i.e. greater than $0$. In this framework, we call \emph{activity} to any pair of positive vectors $\left( \boldsymbol{x},\boldsymbol{y}\right) \in \mathbb{R}_{>0}^m$, where $\boldsymbol{x}\in \mathbb{R}^{m_1}_{>0}$ and $\boldsymbol{y}\in \mathbb{R}^{m_2}_{>0}$. Any $\textrm{DMU}_i \in \mathcal{D}$ has its associated activity $(\boldsymbol{x}_i,\boldsymbol{y}_i)$, where $\boldsymbol{x}_i=(x_{i1},\ldots ,x_{im_1})$ and $\boldsymbol{y}_i=(y_{i1},\ldots ,y_{im_2})$. Therefore, we can identify a DMU with its activity in the same way that a point is identified with its coordinates.

We say that an activity $\left( \boldsymbol{x},\boldsymbol{y} \right) $ \emph{dominates} another activity $\left( \boldsymbol{x}',\boldsymbol{y}' \right) $ if $\boldsymbol{x}\leq \boldsymbol{x}'$ and $\boldsymbol{y}\geq \boldsymbol{y}'$. In this case, $\left( \boldsymbol{x},\boldsymbol{y} \right) $ is considered to be more (or equally) efficient than $\left( \boldsymbol{x}',\boldsymbol{y}' \right) $, because it consumes less (or equal) inputs and produces more (or equal) outputs. The relation ``to be dominated by'' is a partial order on the set of activities. %Moreover, we say that $\left( \boldsymbol{x},\boldsymbol{y} \right) $ \emph{strictly dominates} $\left( \boldsymbol{x}',\boldsymbol{y}' \right) $ if $x_a<x'_a$ for $a=1,\ldots ,m_1$ and $y_b>y'_b$ for $b=1,\ldots ,m_2$.

The \emph{production possibility set} is the set of feasible activities defined by $\mathcal{D}$. Assuming strong disposability of inputs and outputs, and  variable returns to scale (VRS), it is formed by all the activities dominated by convex combinations of DMUs in $\mathcal{D}$. Hence, it is given by
\begin{equation}
\label{eq:p}
\def\arraystretch{1.2}
P = \left\{ \left( \boldsymbol{x},\boldsymbol{y} \right) \in \mathbb{R}_{>0}^m\ \ \Bigg| \
\begin{array}{ll}
\sum_{i=1}^n\lambda _ix_{ig_1}\leq x_{g_1},& g_1=1,\ldots, m_1  \\
\sum_{i=1}^n\lambda _iy_{ig_2}\geq y_{g_2},& g_2=1,\ldots, m_2 \\
\sum_{i=1}^n\lambda _i=1,\quad
\lambda _i\geq 0,& i=1,\ldots ,n 
\end{array} \right\} .
\end{equation}
Constant returns to scale (CRS) can also be considered by removing the convexity condition $\sum_{i=1}^n\lambda _i=1$ from \eqref{eq:p}. However, this convexity condition (and thus the assumption of VRS) is essential for conducting archetypal analysis.

\begin{figure}[htb]
\centering
\includegraphics[width=0.4\textwidth]{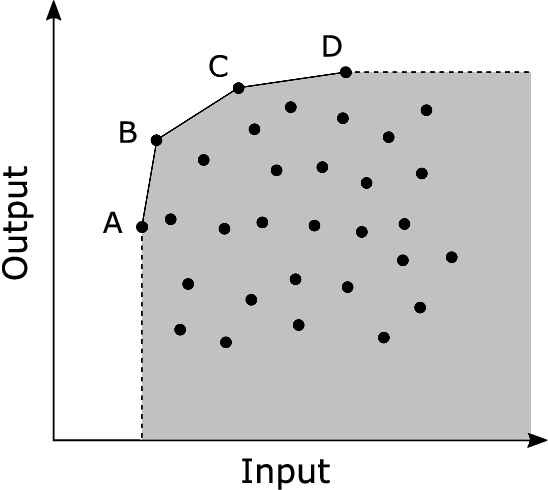}
\caption{Efficient frontier (solid line) and the rest of the frontier (dashed line) of the production possibility set (shaded in grey) for a set of DMUs with one input and one output under VRS. The efficient DMUs are named A, B, C, and D.}
\label{fig:frontier2}
\end{figure}

A DMU (or an activity, in general) is \emph{efficient} if and only if there is no $(\boldsymbol{x},\boldsymbol{y})\in P$, different from the DMU, that dominates the DMU \citep{Koop51}. Hence, a DMU is efficient if and only if there is not another feasible activity in the production possibility set ``better than'' the DMU, in the sense that consumes less (or equal) inputs and produces more (or equal) outputs, with at least one strict inequality. Otherwise, we say that it is \emph{inefficient}.
The set of efficient activities in $P$ is called the \emph{efficient frontier} (also known as \emph{strongly efficient frontier} or \emph{Pareto-Koopmans frontier}) of $P$, and we denote it by $\partial ^{\text{S}}(P)$. It is clear that $\partial ^{\text{S}}(P)$ is in the frontier of $P$, but it does not cover all the frontier (see Figure~\ref{fig:frontier2}).

\subsubsection{Maximal friends sets}
\label{sec:mf}

Given a subset of DMUs, $F\subseteq \mathcal{D}$, it is called \emph{friends} if any convex combination of DMUs in $F$ is efficient.
In \cite{Tone2010} it is proved that $F$ is friends if and only if a convex combination of all DMUs in $F$ is efficient.

A friends set $F$ is said to be \emph{maximal} if it is not contained in any other friends set. Each maximal friends set determines a facet of the efficient frontier. In particular, under the assumption of VRS, a maximal friends set consists of the vertices that define a facet of the efficient frontier. Consequently, the efficient frontier can be fully characterized by the collection of all maximal friends sets (i.e. the vertices of its facets), denoted by $\mathcal{F}$. It follows that, under VRS, any efficient activity in $P$ can be represented as a convex combination of the DMUs belonging to a maximal friends set.
%We show an algorithm for computing $\mathcal{F}$ in Section~\ref{ap:mf}.

%\begin{figure}[htb]
%\centering
%\includegraphics[width=0.35\textwidth]{frontier_gray2.png}
%\caption{Illustration of a part of an efficient frontier under VRS, formed by facets whose vertex sets are maximal friends sets composed by efficient DMUs. Note that the efficient frontier constitutes only a portion of the boundary of the production possibility set and does not cover it entirely.}
%\label{fig:frontier}
%\end{figure}

\section{Methodology}
\label{sec:method}

We now present our semi-supervised archetypal analysis framework, comprising \textit{subarchetypes}, \textit{subarchetypoids}, and several important special cases, including \textit{partitioned archetypes} and \textit{efficient subarchetypes}. Throughout this paper, we use the term ``semi-supervised'' to refer to archetypal analysis in which the construction of archetypal profiles is constrained by a prescribed family of \textit{generating subsets}. These generating subsets may be specified \textit{a priori} by the user, for example to incorporate external knowledge or structural requirements, or they may be determined from intrinsic properties of the data or the underlying problem, as for computing \textit{efficient subarchetypes} in Section~\ref{sec:efsuba}. 

Illustrative toy examples of the proposed methodology are provided in Appendix \ref{sec:ilcomp}.

\subsection{Subarchetypes and subarchetypoids}
\label{sec:suba}

%In this section, we will develop Subarchetype Analysis (subAA) and Subarchetypoid Analysis (subADA).

Let $\mathcal{S}=\left\{ S_p\right\} _{p=1}^r$ be a family of $r$ nonempty subsets of the set of observations $\left\{ \boldsymbol{x}_i\right\} _{i=1}^n$, referred to as \textit{generating subsets}. We note that the union of the generating subsets need not cover the entire set of observations, thereby allowing the framework to accommodate partially labelled datasets. Moreover, the generating subsets are not required to be pairwise disjoint.

Let $\left\{ \boldsymbol{z}_j\right\} _{j=1}^k$ be the set of $k$ \textit{subarchetypes} of $\left\{ \boldsymbol{x}_i\right\} _{i=1}^n$ with respect to $\mathcal{S}$. Each $\boldsymbol{z}_j$ is defined as a convex combination of observations in some generating subset $S_{p(j)}\subseteq \mathcal{S}$, where $p(j)\in \left\{ 1,\ldots ,r\right\} $ for $j=1,\ldots,k$.
Subarchetypes can be written as \eqref{eq:arch}
%\[
%\boldsymbol{z}_j = \sum_{l=1}^n \beta_{jl} \boldsymbol{x}_l,\qquad j=1,\ldots,k,
%\]  
with $\beta_{jl} \geq 0$ and $\sum_{l=1}^n \beta_{jl} = 1$, where $\beta _{jl}=0$ if $\boldsymbol{x}_l \notin S_{p(j)}$.

%Alternativa: Each $\boldsymbol{z}_j$ is defined as a convex combination of observations in some generating subset, i.e. subarchetypes can be written as \eqref{eq:arch}, where
%\[
%\forall j\in \left\{ 1,\ldots,k\right\} \quad \exists p(j)\in \left\{ 1,\ldots ,r\right\} \quad / \quad \beta _{jl}=0 \text{ if } \boldsymbol{x}_l \notin S_{p(j)}.
%\]

Each observation is then approximated as a convex combination of the subarchetypes in the same way as \eqref{eq:alpha}. Finally, the residual sum of squares given by \eqref{eq:rss} is minimized. We will refer to Subarchetype Analysis as subAA.

Note that $p(j)$ specifies the generating subset used to construct the $j$-th subarchetype. Specifically, the $j$-th subarchetype is generated as a convex combination of the elements in the $p(j)$-th generating subset. In the absence of any prior preference, all possible configurations of $p(j)$ must be considered. In general, this results in $\binom{k+r-1}{k}$ distinct configurations, which represents the number of combinations with repetition when distributing $k$ identical elements (subarchetypes) into $r$ distinct categories (generating subsets). Alternatively, prior preferences can be incorporated through $p(j)$ by restricting the set of admissible configurations. For instance, one may impose the constraint that each generating subset generates at most one subarchetype (see Section~\ref{sec:screening}). This excludes all configurations for which there exist $j_1, j_2 \in \{1,\ldots,k\}$ with $j_1 \neq j_2$ such that $p(j_1)=p(j_2)$.

\begin{remark}[\textbf{AA versus subAA}]
\label{rem1}
The distinction between AA and subAA lies in the additional constraints $\beta_{jl}=0$ imposed on the definition of the subarchetypes $\boldsymbol{z}_j$ in \eqref{eq:arch} whenever $\boldsymbol{x}_l \notin S_{p(j)}$. No additional constraints are imposed on the coefficients $\alpha_{ij}$. Since the residual sum of squares to be minimized is given by \eqref{eq:rss}, all observations contribute to the optimization process. Another distinctive feature of subAA is that user-defined lower and/or upper bounds may be imposed on the number of subarchetypes associated with each generating subset.
\end{remark}

\begin{remark}[\textbf{Subarchetypes as a Generalization of Archetypes and Archetypoids}]
\label{rem:gen}
Subarchetypes generalize both archetypes and archetypoids. Specifically, archetypes arise as a special case of subarchetypes when $r=1$ and $S_1=\left\{ \boldsymbol{x}_i\right\} _{i=1}^n$. Conversely, archetypoids correspond to the case $r=n$, with $S_p=\left\{\boldsymbol{x}_p\right\} $ for $p=1,\ldots, n$. Thus, the subarchetype framework encompasses archetypes and archetypoids as two extreme cases.
\end{remark}

Analogously, \textit{subarchetypoids} are defined as archetypoids under subset constraints. In this case, a subarchetypoid $\boldsymbol{z}_j$ is given by \eqref{eq:arch} with $\beta_{jl} \in \left\{ 0,1\right\} $ and $\sum_{l=1}^n \beta_{jl} = 1$, where $\beta _{jl}=0$ if $\boldsymbol{x}_l \notin S_{p(j)}$, and $p(j)$ specifying the generating subset used to construct the $j$-th subarchetypoid. Each observation is approximated by \eqref{eq:alpha}, and the residual sum of squares \eqref{eq:rss} is minimized. As in the case of archetypoids, each subarchetypoid corresponds to a single observation. Moreover, they also generalize archetypoids in the case $r=1$ and $S_1=\left\{ \boldsymbol{x}_i\right\} _{i=1}^n$. We will refer to Subarchetypoid Analysis as subADA.

\begin{remark} [\textbf{ADA versus subADA}]
The distinction between ADA and subADA lies solely in the user-defined lower and/or upper bounds that may be imposed on the number of subarchetypoids associated with each generating subset, provided that every observation belongs to at least one generating subset. If no such bounds are specified, subADA is equivalent to ADA, and the resulting subarchetypoids coincide with the archetypoids.
\end{remark}

An intuitive interpretation of subarchetypes and subarchetypoids is to regard the observations as animals and each generating subset as a group of genetically compatible animals. Subarchetypes can then be interpreted as feasible hybrids within each generating subset, thereby excluding unrealistic combinations analogous to chimeras or crosses between genetically incompatible animals. In contrast, subarchetypoids correspond to actual animals belonging to groups of genetically compatible animals. If no constraints are imposed on the number of subarchetypoids associated with each generating subset, subarchetypoids coincide with archetypoids.

%\begin{remark} In clustering, \textit{centroids} can be viewed as archetypes with $k=1$, whereas \textit{medoids} can be viewed as archetypoids with $k=1$. Accordingly, \textit{subcentroids} may be defined as subarchetypes with $k=1$, without imposing any additional restrictions on the number of subarchetypes associated with each generating subset. Likewise, \textit{submedoids} coincide with \textit{medoids}, provided that no additional restrictions are imposed on the number of subarchetypoids associated with each generating subset.
%\end{remark}

As with AA and ADA, subAA and subADA can be extended to datasets consisting of samples of univariate observed functions (see Section~\ref{sec:faa}), giving rise to Functional Subarchetype Analysis (subFAA) and Functional Subarchetypoid Analysis (subFADA), respectively.

Finally, the number of subarchetypes or subarchetypoids can be selected using criteria analogous to those described in Section~\ref{sec:AAADA} for AA and ADA. The user may specify \emph{a priori} the desired number of representatives generated from each subset, thereby fixing the total number $k$. Otherwise, the elbow criterion can be used to determine $k$, selecting, for each value of $k$, the combination of representatives generated from the subsets that minimizes the $\operatorname{RSS}$.

\subsection{Location of subarchetypes and subarchetypoids}
\label{sec:loc}

The following result about the location of a set of archetypes $\left\{ \boldsymbol{z}_j\right\} _{j=1}^k$ can be found in \citet{cutler1994archetypal}:

\begin{proposition}
\label{prop:cutbre}
Let $C$ be the convex hull of $\left\{ \boldsymbol{x}_i\right\} _{i=1}^n$. Let $B$ be the set of data points on $\partial C$ (the boundary of $C$), and let $N$ be the cardinality of $B$.
\begin{enumerate}
    \item If $k=1$, choosing $\boldsymbol{z}_1$ to be the sample mean minimizes $\operatorname{RSS}$.
    \item If $1<k<N$, there is a set of archetypes $\left\{ \boldsymbol{z}_j\right\} _{j=1}^k$ on $\partial C$ that minimizes $\operatorname{RSS}$.
    \item If $k=N$, choosing $\left\{ \boldsymbol{z}_j\right\} _{j=1}^k=B$ results in $\operatorname{RSS}=0$.
\end{enumerate}
\end{proposition}

The following proposition extends Proposition~\ref{prop:cutbre} to subarchetypes with respect to a collection of generating subsets $\mathcal{S}=\left\{ S_p\right\} _{p=1}^r$, considering all possible configurations:

\begin{proposition}
\label{prop:loc}
Let $C_{\mathcal{S}}$ be the convex hull of $\cup _{p=1}^r S_p$. Let $B_{\mathcal{S}}$ be the set of data points on $\partial C_{\mathcal{S}}$, and let $N_{\mathcal{S}}$ be the cardinality of $B_{\mathcal{S}}$.
Moreover, let $C_p$ be the convex hull of the $p$-th generating subset $S_p$, for $p=1,\ldots ,r$.
\begin{enumerate}
    \item If $k=1$,
    \begin{enumerate}
        \item and the sample mean $\bar{\boldsymbol{x}}$ belongs to $\cup _{p=1}^r C_p$, choosing $\boldsymbol{z}_1=\bar{\boldsymbol{x}}$ minimizes $\operatorname{RSS}$.
        \item and the sample mean $\bar{\boldsymbol{x}}$ does not belong to $\cup _{p=1}^r C_p$, there is a subarchetype $\boldsymbol{z}_1$ on $\partial \cup _{p=1}^r C_p$ that minimizes $\operatorname{RSS}$.
    \end{enumerate}
    \item If $1<k<N_{\mathcal{S}}$, there is a set of subarchetypes $\left\{ \boldsymbol{z}_j\right\} _{j=1}^k$ on $\partial \cup _{p=1}^r C_p$ that minimizes $\operatorname{RSS}$.
    \item If $k=N_{\mathcal{S}}$, choosing $\left\{ \boldsymbol{z}_j\right\} _{j=1}^k=B_{\mathcal{S}}$ yields a minimum of the $\operatorname{RSS}$. Moreover, if $\cup _{p=1}^r S_p=\left\{ \boldsymbol{x}_i\right\} _{i=1}^n$ then $\operatorname{RSS}=0$.
\end{enumerate}    
\end{proposition}

\begin{remark}
    In the case $k=N_{\mathcal{S}}$, the condition $\cup _{p=1}^r S_p=\left\{ \boldsymbol{x}_i\right\} _{i=1}^n$ can be relaxed to $B\subseteq \cup _{p=1}^r S_p$, where $B$ is defined in Proposition~\ref{prop:cutbre}. Since every observation belongs to the convex hull of $B$, this weaker condition is sufficient to ensure that $\operatorname{RSS}=0$.
\end{remark}

\begin{remark}
    Proposition~\ref{prop:loc} generalizes the original result by \citet{cutler1994archetypal} given in Proposition~\ref{prop:cutbre}. Indeed, Proposition~\ref{prop:cutbre} is recovered by taking $r=1$ and $S_1=\{ \boldsymbol{x}_i\} _{i=1}^n$, that is, when subarchetypes coincide with archetypes (see Remark~\ref{rem:gen}). In this special case, $\cup _{p=1}^r S_p=\left\{ \boldsymbol{x}_i\right\} _{i=1}^n$, $C_{\mathcal{S}}=C$, $B_{\mathcal{S}}=B$, $N_{\mathcal{S}}=N$, and $\cup _{p=1}^r C_p=C$.
\end{remark}

Finally, it is worth noting that the location result of Proposition~\ref{prop:loc} does not extend to subarchetypoids. Indeed, just as archetypoids are not necessarily located on the boundary of the convex hull of the observations, subarchetypoids are not necessarily located on the boundary of $\cup _{p=1}^r C_p$. Several counterexamples for archetypoids are given in \citet{vinue2015archetypoids}; the same examples immediately extend to the subarchetypoid setting.

\subsection{Efficient subarchetypes}
\label{sec:efsuba}

Given a set of DMUs, computing archetypes from the set of efficient DMUs does not necessarily produce efficient archetypes. This is because unconstrained convex combinations of efficient DMUs are not guaranteed to remain on the efficient frontier.

The occurrence of inefficient archetypes computed from efficient DMUs is not merely a theoretical possibility, but also arises frequently in practice. For example, in simulations based on 200 DMUs with three inputs and three outputs generated independently from a normal distribution, approximately 30 DMUs are efficient on average. In each simulation, five archetypes were computed from the efficient DMUs. Over 100 simulation runs (500 archetypes in total), about $35\%$ of the resulting archetypes were inefficient. Although this percentage naturally depends on the characteristics of the problem and the data-generating distribution, it shows that obtaining inefficient archetypes from efficient DMUs is not an exceptional situation.

The following proposition provides a sufficient condition for guaranteeing that every subarchetype lies on the efficient frontier.

\begin{proposition}
\label{prop:effsub}
Let $\mathcal{D}$ be a set of DMUs, and let $\mathcal{E}$ denote the corresponding set of efficient DMUs. If the generating subsets are friends sets, then every subarchetype of $\mathcal{E}$ is both feasible and efficient, and therefore lies on the efficient frontier.
\end{proposition}

Assuming VRS, the efficient frontier of $\mathcal{D}$ is composed of facets whose vertex sets are the maximal friends sets, all of which are subsets of $\mathcal{E}$. In this setting, a feasible activity is efficient if and only if it lies on one of these facets, or equivalently, if and only if it can be expressed as a convex combination of the corresponding maximal friends set.
Within the subAA framework, taking maximal friends sets as generating subsets guarantees that the resulting subarchetypes of $\mathcal{E}$ are feasible activities lying on the efficient frontier. Moreover, every point of the efficient frontier belongs to the convex hull of some generating subset, so no region of the efficient frontier is excluded from containing a subarchetype. Such subarchetypes will be referred to as \textit{efficient subarchetypes}.

As an illustration, in Figure~\ref{fig:frontier2}, the set of efficient activities is $\{ A,B,C,D\}$, and the generating subsets are the vertex sets of the facets of the efficient frontier:  $S_1=\{\textrm{A},\textrm{B}\}$, $S_2=\{\textrm{B},\textrm{C}\}$, and $S_3=\{\textrm{C},\textrm{D}\}$. Consequently, every region of the efficient frontier is represented by at least one generating subset, allowing subarchetypes to be located anywhere on the efficient frontier.

\subsection{Partitioned archetypes}
\label{sec:partarch}

A particular case of subarchetypes with a wide range of applications arises when the generating subsets form a partition of the set of observations.

Let $\mathcal{P}=\left\{ P_p\right\} _{p=1}^r$ be a partition of the set of observations $\left\{ \boldsymbol{x}_i\right\} _{i=1}^n$, that is, a collection of nonempty pairwise disjoint subsets whose union is the entire set of observations.
%Each element of $\mathcal{P}$ is a nonempty subset of $\left\{ \boldsymbol{x}_i\right\} _{i=1}^n$ that satisfies $\cup _{p=1}^r P_p = \left\{ \boldsymbol{x}_i\right\} _{i=1}^n$ with $P_{p_1}\cap  P_{p_2} =\emptyset $ for all $p_1,p_2\in\left\{ 1,\ldots ,r\right\} $ and $p_1\neq p_2$.
The corresponding subarchetypes with respect to $\mathcal{P}$ are also referred to as \textit{partitioned archetypes}.

Typically, it is desirable that $\left\{ p(j)\right\} _{j=1}^k=\left\{ 1,\ldots ,r\right\} $, i.e., each element of $\mathcal{P}$ generates at least one partitioned archetype. In this case, the number of possible configurations of $p(j)$ is $\binom{k-1}{r-1}$. For fixed $r$, this quantity is asymptotically of order $k^{r-1}$. However, this condition may be relaxed in the general case without the need to impose any additional restrictions, thereby allowing some elements of the partition not to generate any partitioned archetypes.

Analogously, \textit{partitioned archetypoids} can be defined when subarchetypoids are considered instead of subarchetypes. It is also possible to construct subarchetypoids and  subarchetypes from collections of subsets contained within each element of $\mathcal{P}$. In this case, we can name them \textit{partitioned subarchetypoids} and \textit{partitioned subarchetypes}, respectively.

For example, we could consider an analogy in which observations represent vertebrate animals and the partition elements correspond to major taxonomic groups (mammals, reptiles, birds, fish, and amphibians). Partitioned archetypes within each group identify the most distinctive representatives relative to the rest of the vertebrates. By further considering subsets of genetically compatible animals within each group, the corresponding partitioned subarchetypes avoid unrealistic combinations, in the same way that biological constraints preclude the existence of chimeras. Another example is given in Section~\ref{sec:nba} for NBA players.

%Let us show another example of partitioned subarchetypes. Suppose that we seek to identify the ideal archetypal basketball player for each position (point guard, small forward, shooting guard, centre, etc.). In addition, we require the archetypal player to be efficient with respect to players occupying the same position. We can construct a partition of the set of efficient players (within each position) such that each element of the partition contains the players occupying a given position. For each element of the partition, the generating subsets are defined as the maximal friends sets associated with the efficient frontier of the corresponding position. We would then compute partitioned subarchetypes requiring one partitioned subarchetype generated within each element of the partition, representing the ideal efficient archetypal player for the corresponding position. This example is shown in Section~\ref{sec:nba} for NBA players.

It may also be of interest to apply partitioned archetypal analysis to country-level data, where the partition elements represent groups of countries, such as continents or associations. Similarly, the methodology could be applied to firm-level economic data, with partition elements corresponding, for example, to economic sectors.

\section{Algorithms}
\label{sec:algo}

This section describes the algorithms underlying the proposed methodology. Sections~\ref{sec:subaacomp} and \ref{sec:subadacomp} present the procedures for computing subarchetypes and subarchetypoids, respectively, for a given subset configuration. As in the classical AA and ADA frameworks, these algorithms extend naturally to the functional setting, giving rise to subFAA and subFADA by simply replacing the objective function with the corresponding functional $\operatorname{RSS}$. Section~\ref{sec:cost} analyses the computational cost when all subset configurations are considered and presents a simulation study. Finally, Section~\ref{sec:screening} introduces heuristic screening strategies for reducing the number of candidate subset configurations to be evaluated. %Finally, Section~\ref{ap:mf} describes the computation of maximal friends sets used in the DEA application.

We implemented the algorithms for computing subarchetypes and subarchetypoids in the \texttt{subarchetypes} package for \textsf{R} \citep{R26}. The package allows users either to specify lower and upper bounds on the number of subarchetypal profiles generated by each subset or to provide only the total number of profiles, $k$, in which case the algorithm automatically selects the subset allocation that minimizes the $\operatorname{RSS}$. The computation of maximal friends sets is implemented in the \texttt{deaR} \textsf{R} package \citep{deaR} available at CRAN (\url{https://cran.r-project.org/package=deaR}). The package \texttt{subarchetypes}  and the scripts required to reproduce all experimental results in Section~\ref{sec:results} and Appendix~\ref{sec:ilcomp} and  are available at \url{https://www.uv.es/vbolos/investigacion/subarchetypes.zip}.

\subsection{Subarchetype computation}
\label{sec:subaacomp}

Given a \textit{candidate configuration} specifying how many subarchetypes have to be generated from each generating subset, the subarchetypes are computed as follows. 

We have modified the original alternating minimization algorithm proposed by \citet{cutler1994archetypal}. This algorithm alternates between two constrained optimization steps. First, with the archetypes fixed, each observation is expressed as a convex combination of the archetypes by estimating non-negative coefficients that sum to one. Second, with these coefficients fixed, each archetype is updated as a convex combination of the observed data, again subject to non-negativity and sum-to-one constraints. 

In subAA, this step is modified so that each subarchetype is updated as a convex combination of only the observations assigned to its corresponding generating subset, rather than all observed data. Accordingly, the initialization is also modified. Instead of randomly selecting the initial archetypes, as in the original algorithm, one observation is selected from each predefined generating subset. When these subsets overlap, observations are chosen without replacement so that the same observation is not used to initialize multiple subarchetypes.

These two steps are repeated iteratively until the $\operatorname{RSS}$ converges, yielding a set of subarchetypes that best represent the extreme patterns in the data while providing a convex reconstruction of every observation. The same algorithm applies to both the multivariate and functional settings by simply using the appropriate $\operatorname{RSS}$.

The convergence properties of the proposed algorithm are inherited from those of the alternating optimization algorithm of \citet{cutler1994archetypal}. The only modification consists of restricting each archetype to be a convex combination of the observations belonging to its prescribed generating subset. Consequently, each optimization step remains a convex least-squares problem over a compact convex feasible set. As in the classical algorithm, every iteration decreases (or leaves unchanged) the residual sum of squares, which is bounded below by zero. Therefore, the sequence of objective values is monotone and convergent. The same arguments as those of \citet{cutler1994archetypal} establish convergence of each inner loop to the minimizer of the corresponding restricted convex least-squares problem. As in classical archetypal analysis, however, the alternating optimization procedure is not guaranteed to attain the global minimum, and different initializations may converge to different local minima.

As in $k$-means clustering, AA and subAA are initialized multiple times from different starting points to reduce the risk of becoming trapped in local minima, retaining the solution with the smallest $\operatorname{RSS}$.

\subsection{Subarchetypoid computation}
\label{sec:subadacomp}

Given a candidate configuration specifying how many subarchetypoids have to be generated from each generating subset, the subarchetypoids are computed as follows. 

The algorithm for subADA is identical to that of ADA (see \citet{vinue2015archetypoids} for details), except that the subset constraints are enforced throughout the optimization. In the BUILD phase, the required number of subarchetypoids is selected within each generating subset, without replacement, according to the corresponding restrictions. In the SWAP phase, each subarchetypoid is allowed to swap only with observations belonging to the same generating subset. The algorithm iterates until a sufficiently small value of the $\operatorname{RSS}$ while preserving the prescribed number of subarchetypoids in each subset. As before, the same algorithm applies to both the multivariate and functional settings by simply using the corresponding $\operatorname{RSS}$.

The proposed modifications preserve the main properties of the original ADA algorithm, while reducing the size of the search space. Since each swap is restricted to observations within the corresponding generating subset, the number of candidate exchanges evaluated at each iteration is reduced, potentially leading to faster convergence in practice.

As in subAA, subADA is initialized multiple times from different starting points, and the solution with the smallest $\operatorname{RSS}$ is retained.

Like ADA, subADA can be scaled to large datasets by extending the scalable optimization strategy proposed by \citet{Vinue21}. The only modifications required are to enforce the subset constraints during the BUILD and SWAP phases.

\subsection{Computational cost}
\label{sec:cost}

The computational cost of subAA (and, analogously, of subADA) differs from that of classical AA (and ADA, respectively) mainly because several candidate configurations may have to be evaluated. Throughout this section, the discussion is presented for subAA and AA, but the same arguments apply directly to subADA and ADA. It is therefore useful to distinguish between the cost of solving a single optimization problem and the cost of exploring all candidate configurations.

For a fixed candidate configuration, the computational cost of subAA is essentially the same as that of classical AA. Indeed, once the generating subset assigned to each subarchetype has been fixed, the optimization algorithm is identical to that of AA, except that the coefficients defining each subarchetype are constrained to observations belonging to the corresponding generating subset. Consequently, the computational complexity of one execution of subAA is of the same order as that of AA.

The additional computational effort arises from the need to evaluate multiple candidate configurations. As shown in Section~\ref{sec:suba}, in the absence of any prior preference, the total number of candidate configurations for computing subarchetypes is $\binom{k+r-1}{k}$. For fixed $k$, this quantity grows asymptotically as $\mathcal{O}(r^k)$. More generally, if $T_{\textrm{AA}}$
denotes the computational cost of a single execution of the AA algorithm, then the overall computational cost of subAA is proportional to $N_{\textrm{conf}}T_{\textrm{AA}}$, where $N_{\textrm{conf}}$ is the number of candidate configurations evaluated.

Therefore, the computational bottleneck of subAA is generally not the optimization itself, but the combinatorial growth in the number of candidate configurations. This motivates the screening procedures proposed in Section~\ref{sec:screening}, whose objective is to discard unlikely configurations before running the optimization algorithm.

Another important feature is that the candidate configurations are completely independent. Consequently, each configuration can be evaluated separately, making the proposed methodology embarrassingly parallel and particularly suitable for multi-core and distributed computing environments. This allows substantial reductions in execution time through parallel implementation.

\subsubsection{Simulation study}

To illustrate the computational behaviour of the proposed methodology, we conducted a simulation study focusing on the effect of the number of generating subsets. The remaining parameters (number of observations, dimension of the observations, and number of subarchetypes) were kept fixed, as their influence on the computational cost of AA has already been studied by \citet{eugster2009spider}.

We generated datasets consisting of $n=128$ observations in dimension $m=5$, using a multivariate standard normal distribution. We computed $k=4$ subarchetypes with respect to different numbers of generating subsets, namely $r\in \{1,2,4,8,16\} $. For each value of $r$, the observations were partitioned into generating subsets of the same size.

For each scenario, 50 independent datasets were generated. The complete subAA procedure was then applied to each dataset, evaluating all candidate configurations five times from different initial conditions. For each dataset, the lowest $\operatorname{RSS}$ obtained across all candidate configurations and initial conditions was recorded, together with the total elapsed computation time required for the five runs. All computations were performed using \textsf{R} version 4.5.1 on a computer equipped with an Intel\textsuperscript{\textregistered} Core\textsuperscript{\texttrademark} i9-10900KF CPU @ 3.70 GHz and 64 GB of RAM, running Ubuntu 24.04.4 LTS (Linux kernel 6.8.0-117-generic).

\begin{figure}[htb]
\centering
\includegraphics[width=1\textwidth]{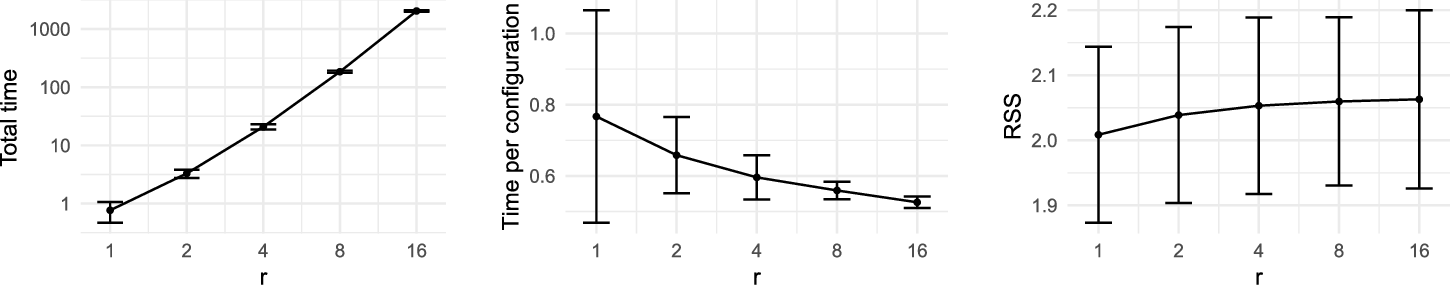}
\caption{Computational performance of subAA for different numbers of generating subsets $r$. The panels show, from left to right, the total computation time (in seconds, on a logarithmic scale), the computation time per candidate configuration, and the resulting $\operatorname{RSS}$. Each point represents the mean over the simulated datasets, with error bars indicating $\pm 1$ standard deviation. The computation times include the five runs from different initial conditions, while the reported $\operatorname{RSS}$ is the lowest value obtained across all candidate configurations and initial conditions.}
\label{fig:sim}
\end{figure}

Figure~\ref{fig:sim} shows the computational results for the different values of the number of generating subsets $r$. The three panels report the total computation time, the computation time per candidate configuration (five runs), and the resulting $\operatorname{RSS}$, respectively. As expected, the total computation time increases substantially with $r$, reflecting the growth in the number of candidate configurations. In contrast, the computation time per configuration decreases as $r$ increases. This behaviour can be attributed to the fact that, for larger values of $r$, each subarchetype is constructed from a more restricted subset of observations, leaving fewer alternatives within each optimization problem and thus facilitating convergence. The decrease in the standard deviation of the computation time per configuration indicates that the computation times become less dispersed as $r$ increases. Regarding the $\operatorname{RSS}$, its increase relative to the classical AA case ($r=1$) is relatively small, and the values tend to stabilize for larger numbers of generating subsets. Thus, in this simulation setting, increasing the number of generating subsets has a comparatively limited effect on the quality of the resulting subarchetypal representation.

% Figure/Table XXX reports the average computation times together with the corresponding numbers of candidate configurations.

% As expected, the computational time increases approximately proportionally to the number of candidate configurations, while the cost of solving each individual optimization problem remains essentially unchanged. This confirms that the principal computational challenge of subAA is the combinatorial growth of the candidate configurations rather than the optimization algorithm itself.

\subsection{Screening candidate configurations}
\label{sec:screening}

Computing subarchetypes requires considering a potentially large number of candidate configurations involving $k$ subarchetypes and $r$ generating subsets, and the same holds for subarchetypoids. As shown in Section~\ref{sec:suba}, the total number of candidate configurations is $\binom{k + r - 1}{k}$, which grows as $\mathcal{O}(r^k)$ for fixed $k$. Consequently, exhaustive evaluation may become computationally prohibitive even for moderate values of $r$.

When the number of candidate configurations becomes prohibitively large, several screening approaches may be used to reduce the number of candidate configurations that need to be evaluated. It should be emphasized, however, that these approaches are heuristic and therefore do not guarantee that the optimal solution will be retained. Nevertheless, they can substantially reduce the computational effort while often yielding solutions with only a small increase in $\operatorname{RSS}$. Examples illustrating the following screening methods are presented in Section~\ref{sec:DEAex}.

\begin{itemize}
\item If the number of generating subsets $r$ is large relative to $k$, we may restrict the search to configurations containing at most one subarchetype per generating subset. Although this reduces the number of candidate configurations to
$\binom{r}{k}$, the asymptotic complexity remains $\mathcal{O}(r^k)$ for fixed $k$.

\item Retain only those generating subsets containing observations that contribute to the classical archetypes or correspond to the classical archetypoids. More generally, additional configurations involving subsets associated with solutions having a low $\operatorname{RSS}$ in the classical archetypal analysis may also be retained, whereas configurations associated with high $\operatorname{RSS}$ values may be discarded.

\item Compute archetypoids using the midpoints of the generating subsets and retain configurations involving those subsets whose midpoints are selected as archetypoids. When seeking \textit{efficient subarchetypes}, where the generating subsets are maximal friends sets (i.e., the vertex sets of the facets of the efficient frontier), configurations involving adjacent facets (namely, facets sharing two or more vertices) may also be considered. The archetypoid analysis may be performed by considering only the midpoints or by considering both the midpoints and the observations (vertices). In the latter case, a partitioned subarchetypoid analysis can be applied to the combined set of midpoints and observations, while constraining the subarchetypoids to be selected from the set of midpoints.
\end{itemize}

In practice, combining several of these strategies often yields substantial reductions in computational effort with only a small increase in the resulting $\operatorname{RSS}$.

% Alternatively, an algorithm similar to that used for ADA  \citep{vinue2015archetypoids} could be developed. Starting from the configuration obtained by computing archetypoids on the generating subset midpoints, the algorithm could iteratively assess whether the solution improves by making local modifications, that is, by moving to nearby configurations.

Finally, although not strictly a screening strategy, it is worth mentioning a practical technique that can drastically reduce the number of evaluated configurations when computing efficient subarchetypes within partition elements. As discussed in Section~\ref{sec:efsuba}, although archetypes computed from efficient DMUs may be inefficient, this situation is relatively uncommon in practice. Consequently, it is often sufficient to introduce the corresponding maximal friends sets as generating subsets only for those partition elements whose subarchetypes are found to be inefficient, while leaving the remaining partition elements undivided. If the resulting subarchetypes are then efficient in all partition elements, there is no need to further refine the remaining partition elements, since doing so would only impose additional constraints and therefore cannot decrease the minimum attainable $\operatorname{RSS}$.

An application of this strategy is presented in Section~\ref{sec:nba}. In that example, only one partition element produces an inefficient subarchetype. In this case, it is sufficient to evaluate only the six candidate configurations associated with the six maximal friends sets of that partition element for all resulting subarchetypes to be efficient. In contrast, introducing the maximal friends sets as generating subsets in every partition element would require evaluating $23\,760$ candidate configurations. This illustrates the substantial computational savings that can be achieved when only a small number of partition elements contain inefficient subarchetypes.

\section{Results and discussion}
\label{sec:results}

\subsection{Motivating applications: efficient subarchetypes}
\label{sec:DEAex}

In this section, we introduce the application that motivates subAA, namely the computation of \textit{efficient subarchetypes} within the framework of DEA (see Section~\ref{sec:efsuba}). It is worth emphasizing that, in this application, the constraints imposed on the subarchetypes arise from the intrinsic geometric structure of the efficient frontier rather than from external expert knowledge.

\subsubsection{Efficient subarchetypes in business data}

To illustrate the proposed methodology, we consider a dataset of 55 decision-making units (DMUs) corresponding to commercial firms from the Valencian Community, recovered from the SABI financial database \cite{sabi}. The firms are medium-sized (employing between 50 and 200 workers) and operated in the wholesale and retail trade sector (Sector G) during 2022. The input variables considered are ``Total Assets'', ``Staff Costs'', and ``Other Costs'', while ``Operating Revenues'' is treated as the output variable. Within this framework, the ``Number of Employees'' is regarded as a non-controllable variable. After conducting DEA  under constant returns to scale (CRS), 12 DMUs were found to be efficient.
%(see Table~\ref{tab:datos_raw}).
Note that we assume CRS because any proportional multiple of a feasible activity can also be considered feasible.

%\begin{table}[ht]
%\centering
%\begin{tabular}{lrrrrr}
%DMU   & Total Assets & Staff Costs & Other Costs & Op. Revenues & N. Employees \\
%\hline
%1  & 21\,175\,101     & 2\,943\,608     & 42\,684\,534    & 47\,618\,378           & 81        \\
%2  & 35\,585\,136     & 3\,566\,899     & 25\,677\,797    & 32\,655\,854           & 57        \\
%3  & 10\,903\,700     & 1\,454\,602     & 29\,578\,241    & 31\,809\,195           & 52        \\
%4  & 5\,397\,210      & 2\,196\,585     & 27\,210\,120    & 30\,228\,823           & 57        \\
%5  & 12\,234\,525     & 4\,012\,901     & 11\,347\,442    & 18\,823\,091           & 197       \\
%6  & 21\,675\,175     & 3\,260\,130     & 12\,033\,604    & 18\,792\,772           & 81        \\
%7  & 14\,299\,600     & 1\,576\,860     & 14\,224\,205    & 18\,579\,445           & 65        \\
%8  & 11\,265\,124     & 2\,083\,528     & 11\,338\,070    & 16\,159\,115           & 51        \\
%9  & 2\,869\,146      & 2\,932\,043     & 6\,975\,082     & 9\,652\,538            & 72        \\
%10 & 1\,589\,629      & 1\,512\,763     & 7\,678\,434     & 9\,564\,337            & 53        \\
%11 & 8\,248\,283      & 2\,581\,488     & 5\,349\,155     & 8\,938\,291            & 60        \\
%12 & 10\,800\,168     & 2\,228\,005     & 4\,401\,847     & 7\,956\,409            & 123      
%\end{tabular}
%\caption{Data from 12 efficient medium-sized companies from Valencian commerce sector in 2022. All data (except ``Number of Employees'') is in euros.}
%\label{tab:datos_raw}
%\end{table}

Our goal is to find efficient subarchetypes from these 12 efficient DMUs. The problem is that the implementation of subAA requires the assumption of variable returns to scale (VRS). However, by virtue of the Theorem of Equivalence for Ratio Variables \citep{Bol2026}, the above efficiency analysis under CRS is equivalent to an efficiency analysis under VRS but expressing all variables as ratios with respect to the non-controllable variable. Specifically, each input and output is normalized by the ``Number of Employees'', thereby allowing the assumption of VRS while preserving the underlying efficiency analysis.

Moreover, to ensure that all variables contribute with comparable relative importance in the computation of archetypes, each ratio variable was normalized by dividing it by its maximum value. This normalization does not affect the efficiency status of the DMUs, as efficiency is invariant to changes in the units of measurement. Subsequently, the output variable was multiplied by $\sqrt{3}$ to compensate for the imbalance between the number of inputs and outputs. Specifically, this adjustment prevents the $\operatorname{RSS}$ used in the computation of archetypes from disproportionately emphasizing the inputs relative to the output, given that the model includes three inputs and only one output. The resulting normalized dataset is presented in Table~\ref{tab:df_eff}.

\begin{table}[ht]
\centering
\begin{tabular}{lrrrr}
DMU    & Input 1 & Input 2 & Input 3 & Output \\
\hline
1  & 0.419        & 0.581       & 0.926       & 1.665              \\
2  & 1.000        & 1.000       & 0.792       & 1.622              \\
3  & 0.336        & 0.447       & 1.000       & 1.732              \\
4  & 0.152        & 0.616       & 0.839       & 1.502              \\
5  & 0.099        & 0.326       & 0.101       & 0.271              \\
6  & 0.429        & 0.643       & 0.261       & 0.657              \\
7  & 0.352        & 0.388       & 0.385       & 0.809              \\
8  & 0.354        & 0.653       & 0.391       & 0.897              \\
9  & 0.064        & 0.651       & 0.170       & 0.380              \\
10 & 0.048        & 0.456       & 0.255       & 0.511              \\
11 & 0.220        & 0.688       & 0.157       & 0.422              \\
12 & 0.141        & 0.289       & 0.063       & 0.183             
\end{tabular}
\caption{Inputs and output considered in the example. All of them are ratio variables, dividing ``Total Assets'', ``Staff Costs'', ``Other Costs'' and ``Operating Revenues'' by the ``Number of Employees''. Moreover, each variable has been divided by its maximum, and the output is multiplied by $\sqrt{3}$ to compensate with respect to the inputs.}
\label{tab:df_eff}
\end{table}

Before computing the efficient subarchetypes, we first calculate the classical archetypes to illustrate that the resulting archetypes are not necessarily efficient. A good choice, according to the elbow-type criterion, is to compute $k=3$ archetypes. The archetypes obtained are $\textrm{DMU}_2$, $0.796\cdot \textrm{DMU}_3+0.204\cdot \textrm{DMU}_4$, and $0.375\cdot\textrm{DMU}_9+0.625\cdot\textrm{DMU}_{12}$ with $\operatorname{RSS}=0.0172$ (see Table~\ref{tab:arch}). The first two archetypes are efficient, but the third is inefficient with an inefficiency score of $0.935$, computed by the input-oriented radial BCC model (values below 1 indicate inefficiency) \citep{Banker1984}. Note that $\textrm{DMU}_9$ and $\textrm{DMU}_{12}$ do not appear together on any facet, see Table~\ref{tab:mf}.

\begin{table}[ht]
\centering
\begin{tabular}{rrrrrrrrrrrrrrrr}
Facet                 & 1 & 2 & 3 & 4 & 5 & 6  & 7 & 8  & 9 & 10 & 11 & 12 & 13 & 14 & 15 \\
\hline
\multirow{4}{*}{DMUs} & 1 & 1 & 1 & 2 & 3 & 3  & 3 & 3  & 4 & 4  & 5  & 5  & 5  & 5  & 5  \\
                      & 2 & 2 & 3 & 3 & 4 & 4  & 4 & 5  & 5 & 5  & 6  & 6  & 6  & 8  & 8  \\
                      & 3 & 4 & 4 & 7 & 5 & 5  & 7 & 7  & 7 & 8  & 7  & 8  & 11 & 9  & 9  \\
                      & 8 & 8 & 8 & 8 & 7 & 10 & 8 & 12 & 8 & 10 & 8  & 11 & 12 & 10 & 11
\end{tabular}
\caption{Facets of the efficient frontier and their corresponding vertex DMUs. Each facet is defined by four vertex DMUs, which constitute a \textit{maximal friends} set.}
\label{tab:mf}
\end{table}

Using the \textit{maximal friends} sets reported in Table~\ref{tab:mf} as generating subsets, the efficient subarchetypes are $\textrm{DMU}_2$, $0.801\cdot \textrm{DMU}_3+0.199\cdot \textrm{DMU}_4$, and $0.716\cdot\textrm{DMU}_5+0.284\cdot\textrm{DMU}_9$, with an $\operatorname{RSS}$ of 0.0186 (see Table~\ref{tab:arch}). All of them are in at least one facet and hence, they are efficient. Specifically, the first efficient subarchetype is in facets 1, 2,  and 3; the second efficient subarchetype is in facets 3, 5, 6, and 7; and the third efficient subarchetype is in facets 14 and 15. As can be seen in Table~\ref{tab:arch}, the efficient subarchetypes closely resemble the archetypes, while achieving only a slightly higher $\operatorname{RSS}$.

In the above computation of efficient subarchetypes, all 680 candidate configurations were considered. Although this does not pose a computational challenge in the present example, the number of candidate configurations may become computationally intractable in more general settings. Several methods for screening candidate configurations can be employed to alleviate this issue (see Section~\ref{sec:screening}). A simple way to reduce the computational burden is to limit the maximum number of subarchetypes per facet to one, which reduces the number of candidate configurations from 680 to 455.

Another of the proposed screening methods requires first computing the archetypoids, motivated by the fact that they tend to be the main contributors to the construction of the efficient subarchetypes. Computing $k=3$ archetypoids yields $\textrm{DMU}_2$, $\textrm{DMU}_3$, and $\textrm{DMU}_5$, with $\operatorname{RSS}=0.0240$ (see Table~\ref{tab:arch}). Thus, we can select candidate configurations involving facets that contain these DMUs. In this case, however, this screening criterion is not effective, since every facet contains at least one of $\textrm{DMU}_2$, $\textrm{DMU}_3$, and $\textrm{DMU}_5$.

On the other hand, the observations contributing to the archetypes are $\textrm{DMU}_2$, $\textrm{DMU}_3$, $\textrm{DMU}_4$, $\textrm{DMU}_9$, and $\textrm{DMU}_{12}$. In this case, two facets do not have any of these DMUs as vertices. Discarding these facets reduces the number of candidate configurations to $455$. If, in addition, we impose that each facet generates at most one subarchetype, the number of candidate configurations is further reduced to $286$.

\begin{table}[ht]
\centering
\begin{tabular}{cl|rrrr}
                                             &                  & Input 1 & Input 2 & Input 3 & Output \\ \hline
\multicolumn{1}{c|}{Archetypes}                        & $\textrm{DMU}_2$ & 1.000   & 1.000   & 0.792   & 1.622  \\
\multicolumn{1}{c|}{}              & $0.796\cdot \textrm{DMU}_3+0.204\cdot \textrm{DMU}_4$               & 0.298   & 0.482   & 0.967   & 1.685  \\
\multicolumn{1}{c|}{$\operatorname{RSS} = 0.0172$}                        & $0.375\cdot\textrm{DMU}_9+0.625\cdot\textrm{DMU}_{12}$               & 0.112   & 0.425   & 0.103   & 0.257 \\ \hline
\multicolumn{1}{c|}{Efficient }                        & $\textrm{DMU}_2$              & 1.000   & 1.000   & 0.792   & 1.622  \\
\multicolumn{1}{c|}{Subarchetypes} & $0.801\cdot \textrm{DMU}_3+0.199\cdot \textrm{DMU}_4$              & 0.299   & 0.481   & 0.968   & 1.686 \\
\multicolumn{1}{c|}{$\operatorname{RSS} = 0.0186$}                        & $0.716\cdot\textrm{DMU}_5+0.284\cdot\textrm{DMU}_9$              & 0.089   & 0.418   & 0.121   & 0.301\\ \hline
\multicolumn{1}{c|}{Archetypoids}                        & $\textrm{DMU}_2$             & 1.000   & 1.000   & 0.792   & 1.622 \\
\multicolumn{1}{c|}{}            & $\textrm{DMU}_3$              & 0.336   & 0.447   & 1.000   & 1.732  \\
\multicolumn{1}{c|}{$\operatorname{RSS}=0.0240$}                        & $\textrm{DMU}_5$              & 0.099   & 0.326   & 0.101   & 0.271
\\ \hline
\end{tabular}
\caption{Results of computing three archetypes, three efficient subarchetypes, and three archetypoids, together with a comparison of their goodness of fit and the coordinates of the obtained representatives.}
\label{tab:arch}
\end{table}

A further screening method proposed in Section~\ref{sec:screening} consists of taking the ``mean activity'' (midpoint) of each facet and performing an archetypoid computation. In this case, the archetypoids correspond to the mean activities of facets 1, 6, and 13.

However, it is also advisable to take the observations into account when identifying the archetypoids among the set of mean activities. To this end, we perform partitioned archetypoid analysis on the union of the mean activities and the observations, using two partition elements: $P_1$, consisting of the mean activities, and $P_2$, consisting of the observations. The partitioned archetypoids are constrained to be selected from $P_1$.
In this case, a similar result is obtained, with the selected mean activities corresponding to facets 1, 3, and 13.

It can be observed that the first efficient subarchetype lies on facets 1 and 3, and the second on facets 3 and 6. Although the third efficient subarchetype does not lie on facet 13, it lies on facet 15, which is adjacent to facet 13 (i.e., the two facets share at least two vertices). These results suggest that filtering candidate configurations using archetypoids computed from the mean activities of the facets can be an effective strategy.

Restricting the candidate configurations to those involving facets 1, 3, or 13, or facets adjacent to them, eliminates only facet 14. This results in $580$ candidate configurations, or $364$ when the maximum number of subarchetypes per facet is restricted to $1$. We may instead focus on configurations containing facets 1, 3, or 13. However, this still leaves a large number of candidate configurations. A more restrictive criterion is to consider the combination of facets 1, 3, and 13 and allow at most one of these facets to be replaced by an adjacent facet, while leaving the remaining facets unchanged. This reduces the number of candidate configurations to $17$, among which is the combination with the minimum $\operatorname{RSS}$ (facets 1, 3, and 15).

Nevertheless, if a sufficiently good solution is acceptable, even if it is not the optimal one, the configuration formed by facets 1, 3, and 13 is the one to select. In this case, restricting the subarchetypes to the combination of facets 1, 3, and 13 yields an $\operatorname{RSS}$ of $0.0193$, which may be regarded as satisfactory, since it is close to the minimum value of $0.0186$. Moreover, using the combination of facets 1, 6, and 13 yields exactly the same solution as the combination of facets 1, 3, and 13, resulting in identical subarchetypes.

\subsubsection{Partitioned efficient subarchetypes in basketball}
\label{sec:nba}

In this example, our aim is to identify the ideal efficient archetypal basketball player for each position. To this end, we compute partitioned subarchetypes from the set of efficient players in each position, with the partition defined according to player position. Within each partition element, the corresponding maximal friends sets are used as generating subsets to ensure that the resulting archetypal players are efficient with respect to players in the same position. The analysis is based on data for all NBA players from the 2025--26 NBA regular season, obtained from \citet{nba}.

The outputs considered are the totals of offensive rebounds (ORB), defensive rebounds (DRB), assists (AST), steals (STL), and points (PTS), while the input is minutes played (MP). In addition, only players with at least 30 games played and averaged at least 10 minutes per game are included in the analysis (the same sample selection as made by \citet{lutz2012,VinEpi17}), i.e. $428$ players in total. Since MP is a non-controllable variable, the Theorem of Equivalence for Ratio Variables \citep{Bol2026} allows us to assume VRS by replacing the original volume variables with the corresponding output-to-MP ratios together with a single input consisting entirely of ones.

There are five different positions: center (C), power forward (PF), point guard (PG), small forward (SF), and shooting guard (SG). After performing DEA separately for each position (applying a radial output-oriented model under VRS), we construct a dataset consisting of the 53 players identified as efficient in their respective positions: 8 C, 10 PF, 13 PG, 8 SF, and 14 SG. 

To perform archetypal analysis on the outputs, we normalize each variable by dividing it by its maximum value and multiplying the result by $100$, thereby expressing all variables as percentages. Since this is merely a linear rescaling, the original values can be recovered at any stage using the maximum values of the ratio variables: $0.214$ (ORB), $0.325$ (DRB), $0.308$ (AST), $0.125$ (STL), and $0.956$ (PTS). As in the previous example, this rescaling does not affect the DEA results. With five player positions, we set $k=5$ for all analyses.

If we compute five \textit{archetypes} without considering player positions, the resulting archetypal profiles combine players from different positions and are therefore not suitable for characterizing players in a specific position. The resulting $\operatorname{RSS}$ of $300.7$ constitutes a lower bound for the $\operatorname{RSS}$ achievable by the subarchetypal analyses considered below, since the subset constraints can only reduce the feasible solution space.

On the other hand, if we define a partition $\mathcal{P}$ according to player position and compute one archetype within each partition element independently (\textit{group-wise archetypes}), we obtain the centroids of the corresponding partition elements. Although this produces a different profile for each position, each profile is computed independently from the others, taking into account only the players in the corresponding position. As a consequence, the resulting profiles are very similar in some variables, such as STL and PTS, and therefore do not adequately capture the differences between playing positions (see Table~\ref{tab:nbares}, Group-wise archetypes). Moreover, they are not efficient, with inefficiency scores of $1.03$ (C), $1.07$ (PF), $1.06$ (PG), $1.04$ (SF), and $1.05$ (SG), computed using the output-oriented radial BCC model (values greater than $1$ indicate inefficiency) \citep{Banker1984}. Although these values are close to $1$, they still indicate that the centroid profiles are inefficient.

\begin{table}[ht]
\centering
\begin{tabular}{rrrrrrrrrrrrr}
                        & \multicolumn{5}{c}{Group-wise archetypes}  &  &                         & \multicolumn{5}{l}{Group-wise efficient subarchetypes}  \\
\multicolumn{1}{r|}{}   & ORB     & DRB    & AST    & STL    & PTS   &  & \multicolumn{1}{r|}{}   & ORB        & DRB      & AST      & STL       & PTS      \\ \cline{1-6} \cline{8-13} 
\multicolumn{1}{r|}{C}  & 61.5    & 78.3   & 40.1   & 35.2   & 57.5  &  & \multicolumn{1}{r|}{C}  & 62.2       & 79.8     & 40.2     & 37.7      & 58.0     \\
\multicolumn{1}{r|}{PF} & 34.0    & 54.3   & 37.1   & 32.6   & 48.8  &  & \multicolumn{1}{r|}{PF} & 34.0       & 54.0     & 37.3     & 36.7      & 49.7     \\
\multicolumn{1}{r|}{PG} & 19.1    & 42.4   & 67.2   & 41.8   & 57.3  &  & \multicolumn{1}{r|}{PG} & 22.8       & 43.2     & 68.9     & 44.0      & 58.4     \\
\multicolumn{1}{r|}{SF} & 25.8    & 50.8   & 50.9   & 35.8   & 66.1  &  & \multicolumn{1}{r|}{SF} & 28.3       & 51.7     & 51.6     & 36.9      & 67.4     \\
\multicolumn{1}{r|}{SG} & 20.8    & 40.2   & 45.3   & 42.7   & 49.8  &  & \multicolumn{1}{r|}{SG} & 21.1       & 43.5     & 46.4     & 44.5      & 50.8     \\
                        &         &        &        &        &       &  &                         &            &          &          &           &          \\
                        & \multicolumn{5}{c}{Partitioned archetypes} &  &                         & \multicolumn{5}{c}{Partitioned efficient subarchetypes} \\
\multicolumn{1}{r|}{}   & ORB     & DRB    & AST    & STL    & PTS   &  & \multicolumn{1}{r|}{}   & ORB        & DRB      & AST      & STL       & PTS      \\ \cline{1-6} \cline{8-13} 
\multicolumn{1}{r|}{C}  & 90.7    & 75.9   & 17.0   & 33.2   & 32.1  &  & \multicolumn{1}{r|}{C}  & 90.2       & 76.2     & 17.1     & 32.7      & 31.9     \\
\multicolumn{1}{r|}{PF} & 16.6    & 52.2   & 59.2   & 28.1   & 31.7  &  & \multicolumn{1}{r|}{PF} & 13.7       & 52.7     & 65.4     & 26.1      & 31.9     \\
\multicolumn{1}{r|}{PG} & 13.2    & 47.5   & 94.4   & 32.2   & 70.1  &  & \multicolumn{1}{r|}{PG} & 12.7       & 45.1     & 94.6     & 33.0      & 71.8     \\
\multicolumn{1}{r|}{SF} & 15.5    & 51.6   & 46.8   & 26.8   & 87.8  &  & \multicolumn{1}{r|}{SF} & 15.5       & 51.6     & 46.9     & 26.5      & 87.7     \\
\multicolumn{1}{r|}{SG} & 16.5    & 26.9   & 18.3   & 100.0  & 37.9  &  & \multicolumn{1}{r|}{SG} & 16.5       & 26.9     & 18.3     & 100.0     & 37.9  
\end{tabular}
\caption{Comparison of the profiles obtained for different player positions using four archetypal analysis approaches: group-wise archetypes, group-wise efficient subarchetypes, partitioned archetypes, and partitioned efficient subarchetypes. All values are expressed as ratios per minute played (MP) and normalized as percentages of the maximum value of the corresponding variable.}
\label{tab:nbares}
\end{table}

To obtain efficient group-wise archetypes, we compute \textit{group-wise efficient sub\-ar\-che\-types}. Specifically, within each partition element, we take the corresponding maximal friends sets as generating subsets, namely 4 for C, 6 for PF, 11 for PG, 5 for SF, and 18 for SG, and compute one efficient subarchetype independently of the other partition elements. The resulting profiles are very similar to the group-wise archetypes, while guaranteeing efficiency in every position (see Table~\ref{tab:nbares} Group-wise efficient subarchetypes). However, as in the previous case, each position is analysed independently, without taking into account the players in the other positions. Consequently, the resulting profiles also fail to adequately capture the differences between playing positions, particularly in variables such as STL and PTS.

If we compute \textit{partitioned archetypes} with respect to $\mathcal{P}$, we obtain an $\operatorname{RSS}$ of $364.7$. Compared with the previous group-wise profiles, partitioned archetypes yield extreme profiles that better characterize the distinctive features of each playing position, since they are computed by jointly considering players from all positions. As a consequence, the resulting profiles are, in a sense, complementary to one another and are therefore more suitable for representing a balanced basketball team (see Table~\ref{tab:nbares}, Partitioned archetypes). However, the PF profile results inefficient, with an inefficiency score of $1.04$, computed using the output-oriented radial BCC model.

To overcome this inefficiency, we use the maximal friends sets associated with the efficient frontier of each position as generating subsets within the corresponding partition elements, and partitioned subarchetypes are computed. Consequently, the resulting archetypal players are guaranteed to be efficient with respect to players in their respective positions. These profiles will be referred to as \textit{partitioned efficient subarchetypes} (see Table~\ref{tab:nbares}, Partitioned efficient subarchetypes).

The partitioned efficient subarchetypes yield an $\operatorname{RSS}$ of $364.9$, which is only slightly larger than that of the partitioned archetypes. Moreover, the resulting profiles are remarkably similar, with only subtle differences (mainly in the PF profile) that are sufficient to guarantee efficiency in every position. Overall, this example illustrates that enforcing efficiency has only a limited impact on the $\operatorname{RSS}$ while guaranteeing that each profile lies on the corresponding efficient frontier.

Finally, it is worth noting that, in this example, it was not necessary to compute partitioned efficient subarchetypes for all $23\,760$ candidate configurations. Since only the PF archetypal profile was inefficient, we considered the maximal friends sets as generating subsets only within the PF partition element. For the remaining positions, the corresponding partition elements were left undivided and therefore acted as single generating subsets. This was sufficient to obtain efficient profiles while requiring the evaluation of only six candidate configurations, corresponding to the six facets of the PF efficient frontier. This illustrates that, in practice, it is often sufficient to refine only those partition elements whose subarchetypes are inefficient. When only a small number of partition elements contain inefficient profiles, this strategy can substantially reduce the computational effort needed to obtain efficient subarchetypes.

\subsection{Applications in human development} 

To illustrate the practical utility of subADA, we consider two applications based on publicly available data from \citet{OWID}. In both cases, countries are grouped according to the official Human Development Index (HDI) \citep{hdi} classification (Very High, High, Medium, and Low Human Development), and the aim is to identify archetypal countries while preserving this grouping structure. The first application considers multivariate societal indicators, whereas the second analyses total fertility rate trajectories as functional data. Together, these examples demonstrate the flexibility of subADA in selecting representative observations from grouped multivariate and functional datasets.

\subsubsection{Societal indicators data} 

The first application considers a multivariate dataset of 166 countries described by four societal indicators. Countries with complete information for the latest available year of each variable were retained, resulting in 65 Very High, 38 High, 39 Medium, and 24 Low HDI countries. The variables are the Human Rights Index (2025) (HRI), Life Expectancy at birth (2023) (LE), Labour Force Participation Rate (2025) (LP), and Scientific and Technical Journal Articles Published per Million People (2022) (SP). All data were obtained from \citet{OWID}.

Because the four indicators are expressed in different units and scales, the analysis was performed using the ranks of each variable rather than their original values. Besides ensuring comparability across variables, this transformation facilitates interpretation, as country performance in these indicators is often communicated through rankings. If there are tied values, we assign them the average of the ranks they would have shared.

Table~\ref{tab:k4} compares the representatives selected by  classical ADA (i.e. without considering the labels) for $k=4$ and subADA by constraining the solution to include one subarchetypoid from each of the four groups.   As expected, ADA achieves a lower reconstruction error ($\operatorname{RSS} = 1\,125.24$) than subADA ($\operatorname{RSS} = 1\,229.74$), since it is not constrained by the HDI grouping. However, ADA selects representatives exclusively from the Very High and Low HDI categories, leaving the High and Medium HDI groups unrepresented. In contrast, since subADA selects one representative from each HDI category, it provides a balanced summary of countries across all levels of human development. In policy applications such as human development, ensuring that all HDI categories are represented is often more valuable for decision-making than achieving a marginal reduction in reconstruction error by concentrating the representatives in only a subset of the groups.

\begin{table}[ht]
\centering
\begin{tabular}{llccccc}
\hline
Method & Country & HDI group & HRI & LE & LP & SP \\
\hline
\multirow{4}{*}{subADA}
& Sweden   & Very high &   2 &  11 &  61 &   7 \\
& China    & High      & 157 &  48 &  62 &  44 \\
& Angola   & Medium    & 107 & 142 &  17 & 166 \\
& Djibouti & Low       & 130 & 137 & 165 & 134 \\
\hline
\multirow{4}{*}{ADA}
& Finland  & Very high & 10.5 &  25 &  98 &   6 \\
& Qatar    & Very high & 135  &  17 &   1 &  33 \\
& Djibouti & Low       & 130  & 137 & 165 & 134 \\
& Niger    & Low       & 102  & 157 &   8 & 161 \\
\hline
\end{tabular}
\caption{Representatives selected by subADA (with one subarchetypoid per group) and classical ADA with $k=4$ for societal indicators data. The variables are expressed as ranks, with tied observations assigned the average of the ranks they would otherwise occupy.}
\label{tab:k4}
\end{table}

Table~\ref{tab:k8} reports the representatives selected when $k=8$ using ADA and subADA, with subADA constrained to select two subarchetypoids from each of the four groups. Again, ADA attains a lower reconstruction error ($\operatorname{RSS} = 268.63$ versus $329.32$ for subADA), but the resulting allocation is unbalanced, with three representatives from the Very High HDI group, three from the Low HDI group, two from the High HDI group, and none from the Medium HDI group. In contrast, as subADA allocates exactly two representatives to each HDI category, it ensures a balanced representation of countries across all development levels.  These results illustrate that subADA can preserve meaningful group representation while incurring only a moderate increase in reconstruction error relative to the unconstrained solution. Notably, even when the number of representatives is doubled from four to eight, ADA still fails to select any representative from the Medium HDI category. In contrast, subADA guarantees balanced representation across all HDI levels while maintaining a competitive reconstruction error. This is particularly appealing in policy applications, where expressing each country as a combination of representatives from all HDI categories provides a more interpretable description than relying on representatives drawn from only a subset of the development levels.

\begin{table}[ht]
\centering
\begin{tabular}{llccccc}
\hline
Method & Country & HDI group & HRI & LE & LP & SP \\
\hline
\multirow{8}{*}{subADA}
& Sweden                & Very high &   2 &  11 &  61 &   7 \\
& Qatar                 & Very high & 135 &  17 &   1 &  33 \\
& Iran                  & High      & 151 &  53 & 155 &  43 \\
& Nicaragua             & High      & 164 &  76 &  48 & 147 \\
& Nigeria               & Medium    &  98 & 166 &   5 & 104 \\
& Sao Tome and Principe & Medium    &  35 & 112 & 166 & 127 \\
& Afghanistan           & Low       & 166 & 136 & 161 & 155 \\
& Niger                 & Low       & 102 & 157 &   8 & 161 \\
\hline
\multirow{8}{*}{ADA}
& Iceland                  & Very high &   6.5 &  13 &  16 &   8 \\
& Italy                    & Very high &  32   &   6 & 145 &  19 \\
& Qatar                    & Very high & 135   &  17 &   1 &  33 \\
& Dominican Republic       & High      &  46.5 &  85 &  57 & 142 \\
& Egypt                    & High      & 152   & 101 & 151 &  67 \\
& Afghanistan              & Low       & 166   & 136 & 161 & 155 \\
& Central African Republic & Low       & 127   & 163 &  18 & 156 \\
& Gambia                   & Low       &  42   & 138 & 149 & 115 \\
\hline
\end{tabular}
\caption{Representatives selected by subADA (with two subarchetypoids per group) and classical ADA with $k=8$ for societal indicators data. The variables are expressed as ranks, with tied observations assigned the average of the ranks they would otherwise occupy.}
\label{tab:k8}
\end{table}

\subsubsection{Functional data: total fertility rate time series} 

The second application considers functional data comprising the total fertility rate trajectories of the same 166 countries analysed in the previous application. The total fertility rate measures the average number of children that a woman would have if she experienced the age-specific fertility rates observed in a given year throughout her reproductive life. The dataset consists of annual observations from 1950 to 2023 and was obtained from \citet{OWID}. As before, the countries are grouped into 65 Very High, 38 High, 39 Medium, and 24 Low HDI countries according to their HDI category.

All functions were expressed using 32 B-spline basis functions of order 4 (cubic splines) over the period 1950--2023, with equally spaced knots, following the same basis specification adopted by \citet{Epifanio2016} in a similar application.

For $k=4$, subFADA (with one subarchetypoid per group) selects Serbia, Thailand, Syria, and Somalia as representatives of the Very High, High, Medium, and Low HDI groups, respectively ($\operatorname{RSS}=8.32$). By contrast, classical FADA (without considering labels) achieves a slightly lower reconstruction error ($\operatorname{RSS}=7.04$) by selecting Latvia, Singapore, Libya, and Afghanistan, resulting in two Very High, one High, one Low, and no Medium HDI representative.

\begin{figure}[ht]
\centering
\begin{tabular}{cc}
\includegraphics[width=1\textwidth]{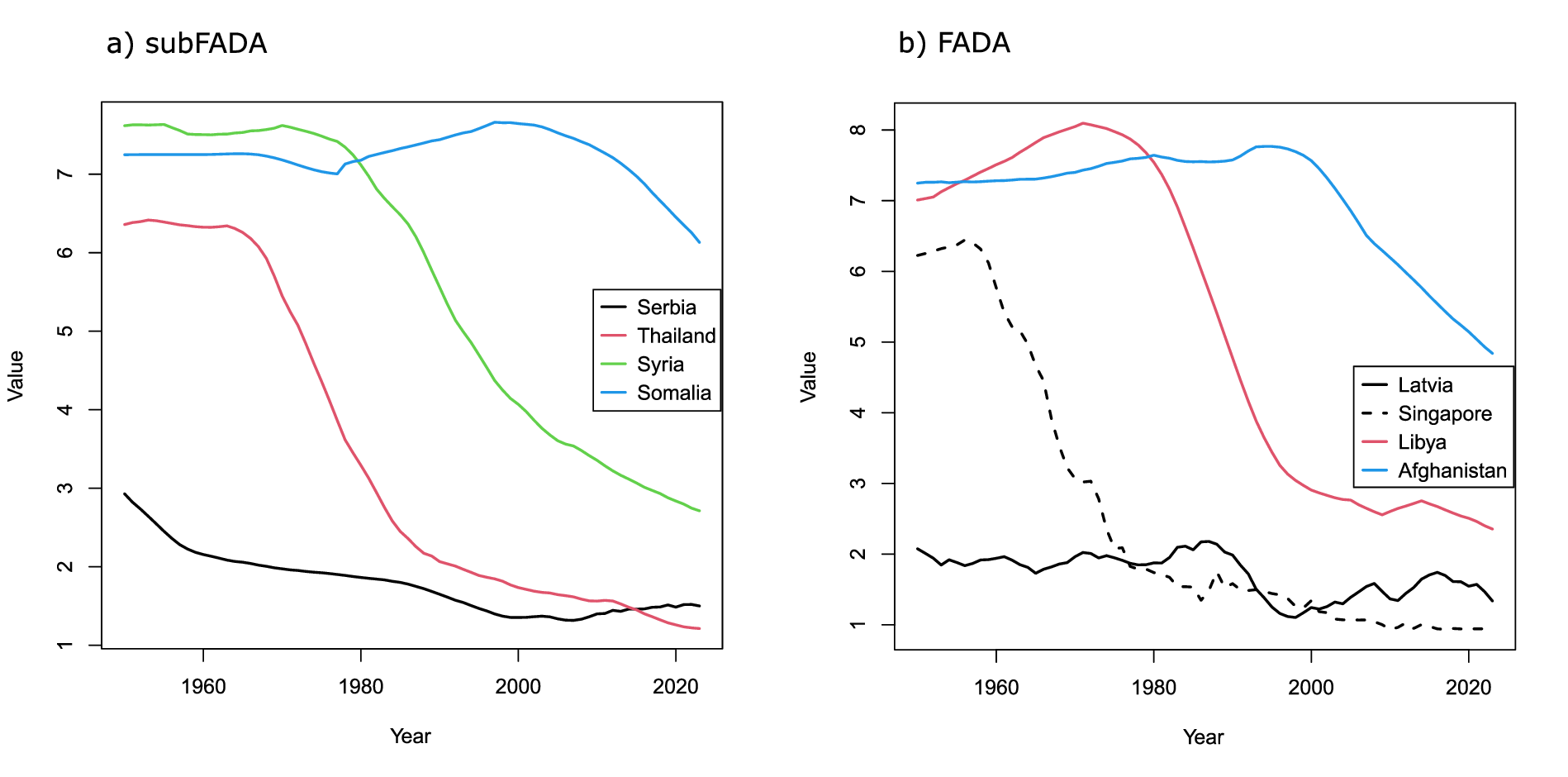}
\end{tabular}
\caption{Fertility rate trajectories of the representatives selected by a) subFADA with one subarchetypoid per group, and b) classical FADA for $k = 4$. Colours indicate the HDI groups: black (Very High), red (High), green (Medium), and blue (Low).}
\label{fig:subFADA_FADA}
\end{figure}

Figure~\ref{fig:subFADA_FADA} shows the fertility trajectories of the selected representatives. Although FADA achieves a slightly lower reconstruction error, two of its representatives exhibit patterns that are similar to those selected by adjacent HDI groups in subFADA. In particular, the trajectory of Singapore (Very High HDI) closely resembles that of Thailand (High HDI), while the trajectory of Libya (High HDI) is similar to that of Syria (Medium HDI). Consequently, FADA allocates two representatives to the Very High HDI group while omitting the Medium HDI group, whereas subFADA preserves one representative for each development level and provides a more balanced characterization of fertility dynamics across HDI categories. From a policy perspective, this is particularly appealing because each country's fertility trajectory is interpreted as a combination of representatives from all HDI categories, rather than from only those groups favoured by the unconstrained optimization.

\section{Conclusions}
\label{sec:conc}

This paper introduces, to the best of our knowledge, the first semi-supervised framework for archetypal analysis. The proposed methodology extends classical AA and ADA by incorporating constraints into the construction of archetypal profiles through subsets of observations. It further generalizes these frameworks and extends the location result for archetypes given by \citet{cutler1994archetypal}. The framework accommodates both user-defined and data-driven generating subsets, allowing external knowledge or intrinsic structural properties of the problem to be incorporated into the analysis, as illustrated by the computation of efficient subarchetypes within the framework of DEA. The methodology has been developed for both multivariate and functional data, and the corresponding algorithms have been implemented in the \texttt{subarchetypes} \textsf{R} package.

The experimental results illustrate the flexibility of the proposed framework across both user-defined and data-driven settings. In the DEA applications, the methodology provides a natural way to compute efficient subarchetypes by restricting the admissible subsets to the vertex sets of the efficient frontier facets, called maximal friends sets. In the human development applications, the subgroup-constrained approaches produce balanced and interpretable representatives, ensuring that all predefined groups are represented while jointly minimizing the global reconstruction error. Comparisons with constrained clustering, group-wise archetypal analysis, and the classical unconstrained approach further highlight the advantages of optimizing the global reconstruction error under subset constraints rather than analysing each group independently or ignoring the available prior information.

Several directions for future research remain open. An immediate extension concerns the computation of efficient subarchetypes under alternative DEA technologies, including constant (CRS), non-increasing (NIRS), non-decreasing (NDRS), and generalized (GRS) returns to scale \citep{Cooper2007}. Since these technologies modify the admissible combinations of observations through different returns-to-scale assumptions, extending the proposed framework would require new optimization procedures.

Another important line of research is the development of scalable implementations for large datasets by adapting the computational strategies proposed for scalable AA and ADA \citet{alcacer2025survey}. Moreover, an algorithm for computing subarchetypes could be developed without requiring the exploration of all candidate configurations, following an approach similar to that of the current archetypoid algorithm. Starting from an initial configuration, the algorithm could iteratively move to ``nearby'' configurations that improve the $\operatorname{RSS}$ until a sufficiently good solution is obtained.

Furthermore, methodological extensions for subarchetypal analysis can be carried out. These include robust formulations based on alternative loss functions \citep{Moliner2019}, kernel formulations to capture nonlinear relationships in the data \citep{morup2012archetypal}, as well as subarchetypal analysis for other data types, including shapes \citep{EpiIbSi17,plos20,epifanio2023archetypal}, binary data \citep{sort20}, ordinal data \citep{FERNANDEZ2021281}, datasets with missing values \citep{doi:10.1080/00031305.2018.1545700}, asymmetric dissimilarity data \citep{VinEpi17,alcacer2025multidimensional}, and extending to biarchetypes \citep{alcacerieee24}, where subsets of both observations and variables will be considered. Another promising direction is the incorporation of fairness into subarchetypal analysis. While fairness has recently been addressed in AA through constraints on the representation coefficients \citep{alcacer2025incorporating}, its integration with subset-constrained archetypal analysis remains an open research question.

Finally, we expect the proposed framework to find useful applications in many domains where prior structural information is naturally available.

\section*{Declaration of generative AI and AI-assisted technologies in the
writing process} During the preparation of this work the authors used OpenAI’s ChatGPT in order to improve language and readability. After using this tool, the authors reviewed and edited the content as needed and take full responsibility for the content of the published article.

\section*{Acknowledgments} We thank Morten M{\o}rup for his constructive feedback and insightful suggestions.

\section*{Funding} This work was partially supported by the Spanish Ministry of Science and Innovation PID2023-153128NB-I00, PID2022-141699NB-I00 and PID2020-118763GA-I00 and Generalitat Valenciana CIPROM/2023/66.

\section*{Declaration of competing interest}  The authors have declared that no competing interests exist.

\section*{Data and code availability} The package \texttt{subarchetypes}  and the scripts with data required to reproduce all experimental results in Section~\ref{sec:results} and Appendix~\ref{sec:ilcomp}  are available at \url{https://www.uv.es/vbolos/investigacion/subarchetypes.zip}.

\appendix

\section{Proofs}
\label{sec:proofs}

\begin{proof}[Proof of Proposition \ref{prop:loc}]

In each case, the proposed subarchetypes satisfy the subset constraints, since each of them can be expressed as a convex combination of observations belonging to a common generating subset. It therefore remains to prove their optimality with respect to the $\operatorname{RSS}$.
\begin{enumerate}

    \item $k=1$.
    \begin{enumerate}
    
        \item The sample mean $\bar{\boldsymbol{x}}$ is the unconstrained minimizer of the $\operatorname{RSS}$.
        
        \item It is clear that the subarchetype $\boldsymbol{z}_1$ belongs to $\cup _{p=1}^r C_p$ since it belongs to one of the sets $C_p$. Moreover, an optimal subarchetype exists because $\cup _{p=1}^r C_p$ is compact and the $\operatorname{RSS}$ is continuous.
        
        Suppose that $\boldsymbol{z}_1$ is an interior point of $\cup _{p=1}^r C_p$. Let
        \[
        \boldsymbol{z}(t)=\bar{\boldsymbol{x}}+t(\boldsymbol{z}_1-\bar{\boldsymbol{x}}),\qquad 0<t<1,
        \]
        and define
        \[
        \operatorname{RSS}(t)=\sum _{i=1}^n\| \boldsymbol{x}_i-\boldsymbol{z}(t)\| ^2.
        \]
        Since $\boldsymbol{z}_1$ is an interior point of $\cup _{p=1}^r C_p$, there exists $\varepsilon >0$ such that $\boldsymbol{z}(t)\in \cup _{p=1}^r C_p$ for all $1-\varepsilon<t<1$. Moreover, a direct calculation shows that
        \[
        \operatorname{RSS}(t)=\operatorname{RSS}(0)+nt^2\| \boldsymbol{z}_1-\bar{\boldsymbol{x}}\| ^2,
        \]
        and hence $\operatorname{RSS}(t)$ is strictly increasing for $0<t<1$ because $\boldsymbol{z}_1\neq \bar{\boldsymbol{x}}$. Therefore, $\boldsymbol{z}(t)$ is feasible and $\operatorname{RSS}(t)<\operatorname{RSS}(1)$ for all $1-\varepsilon<t<1$, contradicting the optimality of $\boldsymbol{z}_1$.
    \end{enumerate}
    
    \item $1<k<N_{\mathcal{S}}$. It is clear that each subarchetype belongs to $\cup _{p=1}^r C_p$ since it belongs to one of the sets $C_p$. Moreover, an optimal set of subarchetypes exists because $\cup _{p=1}^r C_p$ is compact and the $\operatorname{RSS}$ is continuous.
    
    Suppose without loss of generality that $\boldsymbol{z}_1$ is an interior point of $\cup _{p=1}^r C_p$. Let
    \[
    \boldsymbol{z}(t)=\boldsymbol{z}_j+t(\boldsymbol{z}_1-\boldsymbol{z}_j),\qquad t>1,
    \]
    for some $j\neq 1$, and choose $t>1$ such that $\boldsymbol{z}(t)\in \partial \cup _{p=1}^r C_p$. Such a value of $t$ exists because $\cup _{p=1}^r C_p$ is compact and $\boldsymbol{z}_1$ is an interior point. For $\boldsymbol{z}_1,\ldots ,\boldsymbol{z}_k$ fixed, $\operatorname{RSS}$ is minimized with respect to the coefficients $\alpha_{ij}$ by choosing $\sum _{j=1}^k \alpha _{ij}\boldsymbol{z}_j$ to be the point in the convex hull of $\{ \boldsymbol{z}_1,\ldots ,\boldsymbol{z}_k\} $ closest (in the Euclidean metric) to $\boldsymbol{x}_i$, for all $i=1,\ldots ,n$.
    
    Since the convex hull of $\{ \boldsymbol{z}(t),\boldsymbol{z}_2,\ldots ,\boldsymbol{z}_k\} $ contains the convex hull of $\{ \boldsymbol{z}_1,\ldots ,\boldsymbol{z}_k\} $, minimizing the $\operatorname{RSS}$ with respect to the coefficients $\alpha _{ij}$ over the former cannot yield a larger value than minimizing it over the latter. If the new $\operatorname{RSS}$ is strictly smaller, we contradict the optimality of $\left\{ \boldsymbol{z}_j\right\} _{j=1}^k$. Otherwise, the same minimum is attained with $\boldsymbol{z}(t)$ lying on the boundary of $\cup _{p=1}^r C_p$. Hence, there always exists an optimal solution in which $\boldsymbol{z}_1$ belongs to $\partial \cup _{p=1}^r C_p$.

    \item $k=N_{\mathcal{S}}$. We may choose the subarchetypes to be precisely the vertices of $C_{\mathcal{S}}$, that is, $\left\{ \boldsymbol{z}_j\right\} _{j=1}^k=B_{\mathcal{S}}$. Hence, the convex hull of $\left\{ \boldsymbol{z}_j\right\} _{j=1}^k$ is $C_{\mathcal{S}}$, which is the largest feasible convex set obtainable under the generating-subset constraints. Therefore, no other feasible choice of subarchetypes can generate a larger feasible region, and consequently no smaller value of the $\operatorname{RSS}$ can be achieved.

    Moreover, if $\cup _{p=1}^r S_p=\left\{ \boldsymbol{x}_i\right\} _{i=1}^n$ then every observation belongs to $C_{\mathcal{S}}$. So, each observation is represented exactly by a convex combination of the subarchetypes, and therefore $\operatorname{RSS}=0$.

\end{enumerate}

\end{proof}

\begin{proof}[Proof of Proposition \ref{prop:effsub}]
Every subarchetype of $\mathcal{E}$ is a convex combination of DMUs belonging to a generating subset, which is a friends set. Since any convex combination of DMUs in a friends set is efficient (see Section~\ref{sec:mf}), every subarchetype of $\mathcal{E}$ is efficient. Moreover, regardless of the returns to scale, any convex combination of DMUs is feasible. Hence, every subarchetype of $\mathcal{E}$ is also feasible.
\end{proof}

\section{Illustrative comparisons}
\label{sec:ilcomp}

To illustrate the behaviour of the proposed subAA and subADA, we consider several toy examples. We compare the new methodologies with three alternative approaches. Firstly, we consider constrained clustering. Secondly, we compare subADA with a group-wise ADA approach, which serves as an alternative baseline where archetypoids are estimated independently within each generating subset. Thirdly, we compare the proposed subADA with classical ADA, which ignores group labels. These qualitative examples highlight the conceptual differences between these methods and provide intuition about the effect of incorporating group information into the archetypal estimation process.

\subsection{Constrained clustering versus subAA}

To the best of our knowledge, there is no method in the literature that is directly comparable to semi-supervised archetypal analysis. The closest related approaches are found in constrained clustering \citep{GonzalezAlmagro2025}, which are semi-supervised clustering methods, where partial label information or pairwise constraints are used to guide the partitioning of the data. However, semi-supervised clustering techniques do not retain the convex geometric structure that characterizes archetypal representations, since they primarily focus on cluster assignment rather than on learning interpretable extremal representations of the data space, as in subAA. In other words, whereas constrained clustering focuses on restricting how observations are assigned to clusters, subAA constrains how archetypal representatives are constructed. 

If we express $k$-means in a matrix factorization framework as AA was defined in Section~\ref{sec:AAADA}, the matrix $\boldsymbol{Z}_{k \times m} = (\boldsymbol{\alpha}^\prime \boldsymbol{\alpha})^{-1}\boldsymbol{\alpha}^\prime \boldsymbol{X}_{n \times m}$ contains the centroids of each of the $k$ groups that partition the data set, while the constraints are  $\sum_{j=1}^k \alpha_{ij} = 1$ with $\alpha_{ij} \in \{0,1\}$ for $i=1,\ldots,n$. In constrained clustering, the restrictions are imposed on the $\alpha$ values. However, in subAA, the restrictions are not placed on the $\alpha$ values, but on the $\beta$ values, and consequently on $\boldsymbol{Z}_{k \times m}$ (see Remark~\ref{rem1}).

The distinction between these two approaches is illustrated using data from the World Happiness Report dataset \citep{helliwell2019worlddataset}. In this example, we consider two variables, \texttt{GDP.per.capita} and \texttt{Healthy.life.expectancy}, together with the happiness score, which ranges from 0 to 10. To provide partial class information, we label 10\% of the countries based on their happiness scores, assigning the 5\% with the highest scores to the happy class and the 5\% with the lowest scores to the unhappy class. This strategy ensures a clear distinction between the two classes while avoiding the arbitrary choice of a threshold for converting the continuous happiness score into categorical labels \citep{hand2006classifier}. The remaining 90\% of the countries are left unlabelled.

\begin{figure}[htb]
\centering
\includegraphics[width=0.5\textwidth]{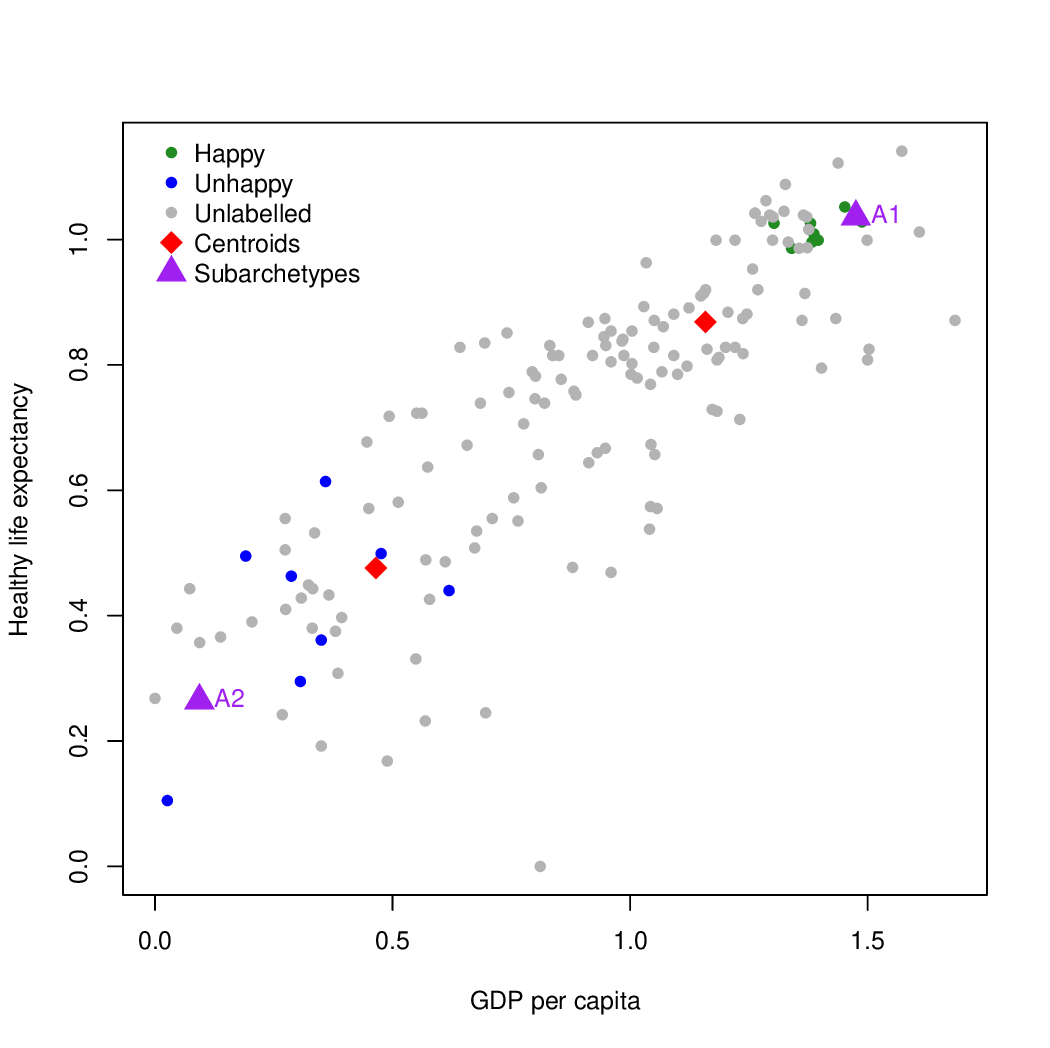}
\caption{World Happiness Dataset with centroids of constrained $k$-means and subarchetypes.}
\label{fig:happy}
\end{figure}

In Figure~\ref{fig:happy}, we visualize the distribution of happy, unhappy and unlabelled countries in the space defined by the two selected variables. Additionally, the figure shows the centroids obtained from constrained $k$-means \citep{conclust,wagstaff2001constrained} and the subarchetypes identified with $k=2$, considering as generating subsets the two sets of happy and unhappy countries and 
computing one subarchetype per class. These subarchetypes represent extreme prototypes of the data distribution, corresponding here to Utopia (highest happiness) and Dystopia (lowest happiness), commonly used in sociology as a benchmark. In contrast, clustering centroids represent average profiles located near the centre of the data cloud, resulting in less distinctive characteristics.

Another difference is that constrained $k$-means only provides group membership, whereas subAA expresses each observation as a convex combination of the identified subarchetypes, thereby revealing its relative proximity to Utopia and Dystopia. For example, constrained clustering assigns both a country with GDP per capita and healthy life expectancy values of 0.657 and 0.672, respectively, and another country with corresponding values of 0.204 and 0.390 to the cluster of unhappy countries. Moreover, both countries are equidistant from their corresponding cluster centroid. Despite this, subAA provides a more informative representation. Specifically, the first country is expressed as 44\% Utopia and 56\% Dystopia, whereas the second is represented as 10\% Utopia and 90\% Dystopia. Hence, subAA distinguishes the profiles of these two countries, whereas constrained clustering provides the same cluster assignment for both.

Furthermore, unlike constrained $k$-means, subAA allows the user to specify an arbitrary number of subarchetypes for each group. In contrast, constrained $k$-means is limited to a single representative (centroid) per group.

\subsection{Group-wise ADA versus subADA}

To illustrate the difference between group-wise ADA  and subADA, we consider a simple two-dimensional toy example. Here, group-wise ADA serves as an alternative baseline in which archetypoids are estimated separately for each generating subset, with each optimization minimizing the $\operatorname{RSS}$ corresponding exclusively to the observations within that subset.

The dataset consists of ten observations representing cities from two countries $A$ and $B$, described by average annual temperature and annual precipitation. A partition $\mathcal{P}=\{P_p\}_{p=1}^2$ is defined, where $P_1$ contains cities from country $A$ and $P_2$ contains cities from country $B$. The two countries partially overlap in the feature space, with one city from each country sharing the same coordinates $(15,700)$. We compare the representatives selected by applying ADA independently within each country (and two archetypoids per country) with those obtained by subADA using $P_1$ and $P_2$ as the generating subsets (and two subarchetypoids per subset).

\begin{figure}[htb]
\centering
\includegraphics[width=0.6\textwidth]{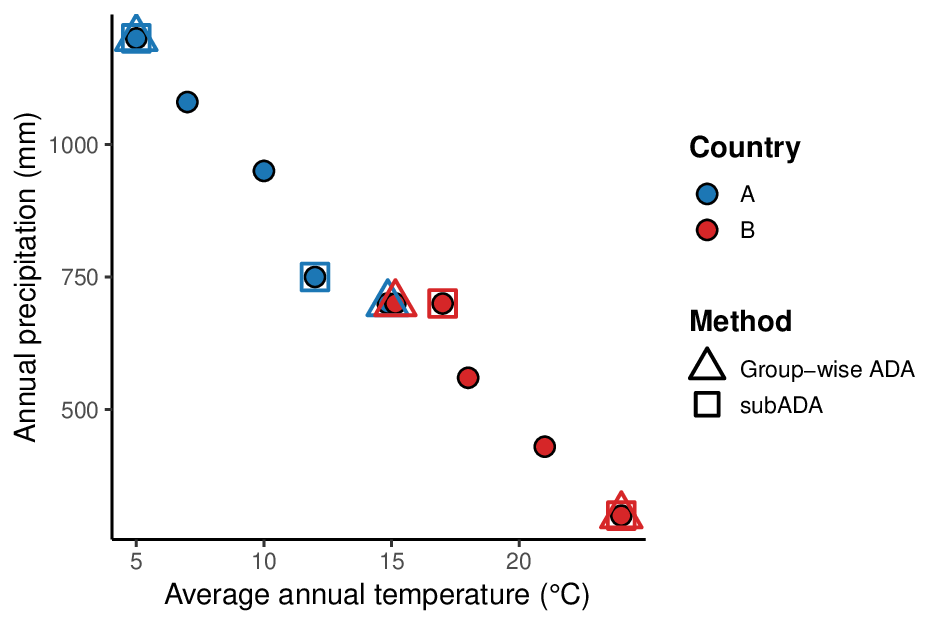}
\caption{Toy example illustrating the difference between group-wise ADA and subADA. Two cities, one from each country, share the same coordinates. They are displayed with a slight offset for visualization purposes.}
\label{fig:subADA_ADAg}
\end{figure}

Figure~\ref{fig:subADA_ADAg} compares the representatives selected by the two approaches.  Because two countries overlap in the feature space, group-wise ADA selects two archetypoids located at the same point, leading to redundant representatives. By jointly minimizing the global reconstruction error while enforcing two archetypoids per country, subADA avoids this redundancy and selects a more informative set of representatives that better captures the overall distribution of the data.

The advantage of subADA is even more apparent when only one representative is selected per country. While group-wise ADA independently selects the medoid of each country, located at $(10,950)$ and $(18,560)$, subADA selects the cities at $(5,1200)$ and $(24,300)$, which represent the two extremes of the complete dataset. As a result, subADA achieves a more informative representation of the global data structure, whereas the group-wise solution is optimal only within each individual country.

\subsection{ADA versus subADA}

\begin{figure}[htb]
\centering
\includegraphics[width=0.6\textwidth]{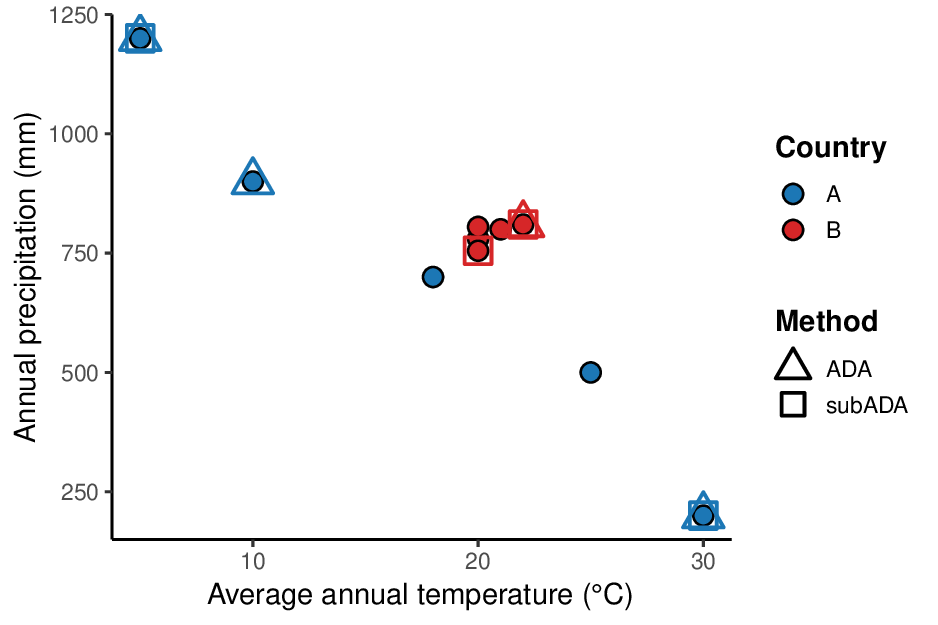}
\caption{Toy example illustrating the difference between ADA and subADA.}
\label{fig:subADA_ADA}
\end{figure}

To illustrate the differences between classical ADA (without considering groups) and subADA, we consider a simple two-dimensional toy example (see Figure~\ref{fig:subADA_ADA}). The dataset consists of ten observations representing cities from two countries, described by two variables: average annual temperature and annual precipitation. As in the previous example, a partition $\mathcal{P}=\{P_p\}_{p=1}^2$ is defined, where $P_1$ contains cities from country $A$ and $P_2$ contains cities from country $B$. Country $A$ exhibits substantially greater variability than country $B$, whose observations are concentrated within a relatively small region of the feature space. 

We compare the representatives selected by ADA with $k = 4$ archetypoids, with those obtained by subADA using $P_1$ and $P_2$ as the generating subsets (and two subarchetypoids per subset). This example demonstrates that, although both approaches use the same number of representatives, the global solution given by ADA tends to allocate more archetypoids to the more heterogeneous country, whereas the subgroup-based approach given by subADA guarantees balanced representation of both countries while still capturing within-group variability.

\bibliography{ref}
\bibliographystyle{abbrvnat}

\end{document}